\documentclass[fleqn,usenatbib]{mnras}

\usepackage{newtxtext,newtxmath}
\usepackage[T1]{fontenc}

\DeclareRobustCommand{\VAN}[3]{#2}
\let\VANthebibliography\thebibliography
\def\thebibliography{\DeclareRobustCommand{\VAN}[3]{##3}\VANthebibliography}

\usepackage{graphicx}	% Including figure files
\usepackage{amsmath}	% Advanced maths commands
\usepackage{subcaption}
\usepackage{graphicx}
\usepackage{longtable}
\usepackage{tabularx}
\usepackage{tikz}
\title[PAHs in the Atmosphere of WASP-6~b]{Detectability of Polycyclic Aromatic Hydrocarbons in the Atmosphere of WASP-6~b with \textit{JWST} NIRSpec PRISM}

\author[Grübel et al.]{
Fabian Grübel,$^{1,2}$\thanks{E-mail: fgruebel@usm.uni-muenchen.de}
Karan Molaverdikhani,$^{1,2}$
Barbara Ercolano,$^{1,2}$
Christian Rab,$^{1,3}$
\newauthor\ Oliver Trapp,$^{2,4}$
Dwaipayan Dubey$^{1,2}$
and Rosa Arenales-Lope$^{1,2}$
\\
$^{1}$Universitäts-Sternwarte, Fakultät für Physik, Ludwig-Maximilians-Universität München, Scheinerstr. 1, D-81679 München, Germany\\
$^{2}$Exzellenzcluster `Origins’, Boltzmannstr. 2, D-85748 Garching, Germany\\
$^{3}$Max-Planck-Institut für Extraterrestrische Physik, Giessenbachstr. 1, 85748 Garching, Germany\\
$^{4}$Fakultät für Chemie und Pharmazie, Ludwig-Maximilians-Universität München, Butenandtstr. 5–13, 81377 München, Germany
}

\date{Accepted XXX. Received YYY; in original form ZZZ}

\pubyear{2024}

\begin{document}
\label{firstpage}
\pagerange{\pageref{firstpage}--\pageref{lastpage}}
\maketitle

\begin{abstract}
Polycyclic Aromatic Hydrocarbons (PAHs) have been detected throughout the universe where they play essential roles in the evolution of their environments. For example, they are believed to affect atmospheric loss rates of close-in planets and might contribute to the pre-biotic chemistry and emergence of life. Despite their importance, the study of PAHs in exoplanet atmospheres has been limited. We aim to evaluate the possibility of detecting PAHs on exoplanets considering future observations using \textit{JWST}'s NIRSpec PRISM mode. The hot Saturn WASP-6~b shows properties that are consistent with a potential PAH presence and is thus used as a case study for this work. Here, we compare the likelihoods of various synthetic haze species and their combinations with the influence of PAHs on the transmission spectrum of WASP-6~b. This is possible by applying the atmospheric retrieval code petitRADTRANS to a collection of data from previous observations. Subsequently, by exploring synthetic, single transit \textit{JWST} spectra of this planet that include PAHs, we assess if these molecules can be detected in the near future. Previous observations support the presence of cloud/haze species in the spectrum of WASP-6~b. While this may include PAHs, the current data do not confirm their existence unambiguously. Our research suggests that utilizing the JWST for future observations could lead to a notable advancement in the study of PAHs. Employing this telescope, we find that a PAH abundance of approximately 0.1 per cent of the ISM value could be robustly detectable.
\end{abstract}


%%%%%%%%%%%%%%%%%%%%%%%%%%%%%%%%%%%%%%%%%%%%%%%%%%

%%%%%%%%%%%%%%%%% BODY OF PAPER %%%%%%%%%%%%%%%%%%

\section{Introduction}
\label{Introduction}
We all have produced PAHs at home when burning food. If not, we have most likely inhaled them in polluted cities or when someone smokes. This distinct group of molecules is defined by their chemical constituents, including multiple benzene rings. Thus, they come in all shapes and sizes ranging from a few ten to hundreds of carbon atoms \citep[for examples, refer to NASA Ames PAH database:][]{2014ApJS..211....8B,2018ApJS..234...32B, 2020ApJS..251...22M}. Their diversity is sheer limitless as they additionally tend to form molecular complexes, so-called PAH clusters, with sizes up to tens of thousands of single molecules depending on the environment they are thriving in \citep{2021A&A...653A..21L, 2023A&A...674A.200L}.

PAHs have been found all over the universe thanks to their unique absorption cross-sections with multiple pronounced features. The corresponding infrared transitions are generated by C-H and C-C stretching and bending modes \citep[e.g.][]{2007ApJ...657..810D} which are shared by most PAH species regardless of the size. As a consequence, scientists were able to detect these features beyond the confines of our home planet e.g. on Saturn's hazy moon Titan using data from the space probe Cassini \citep{Dinelli2013,2013ApJ...770..132L}. Accordingly, the upper atmosphere harbours a PAH abundance of about $(2-3) \times 10^{4}$ particles per cubic centimetre with observed sizes ranging from 9 to 96 carbon atoms. Their presence in the ISM is well-established with a relative number density of approximately $3\times10^{-7}$ respective to hydrogen nuclei \citep{2008ARA&A..46..289T}. Furthermore, PAHs are known to be part of all types of protoplanetary discs \citep[e.g.][]{2006A&A...459..545G,2010ApJ...718..558A, 2021A&A...652A.133V}. Their spectral trademarks are in fact so prominent that PAHs can even be detected in extragalactic sources with the \textit{James Webb Space Telescope} (\textit{JWST}) \citep{2023ApJ...944L..11C,2023ApJ...944L...7S}.\\ 

However, not only are these molecules known for their unique infrared features, their existence usually does not stay without consequence as they alter and shape their environments. For example, they efficiently turn UV photons into heat \citep{1994ApJ...427..822B}. Thus, they play a crucial role in heating the ISM within photo-dissociation regions \citep[e.g.][and references therein]{2022ARA&A..60..247W}. 
\begin{table}
\caption{Relevant parameters of the WASP-6 system used in the atmospheric retrievals and the creation of the synthetic datasets. Parameters from top to bottom: Radius, effective temperature, metallicity, surface gravity, J magnitude (2MASS), radius, mass, transit duration.}
\label{tab:W6b}
\centering
\begin{tabular*}{0.9\columnwidth}{@{\extracolsep{\fill}}lcc@{}}
\hline
\hline
Parameter & Value & References \\
\hline
\noalign{\smallskip}
& WASP-6 & \\
\noalign{\smallskip}
\hline
$R/R_{\sun}$ & 0.870 & 1     \\
$T_{\mathrm{eff}}[K]$ & 5450 & 1     \\           
$\log[\mathrm{Fe/H}]$ & -0.20 & 1     \\
$\log(g)[\mathrm{cgs}]$ & 4.58 & 2    \\
$m_J[\mathrm{mag}]$ & 10.769 & 3 \\  
\hline
\noalign{\smallskip}
\multicolumn{3}{c}{WASP-6~b} \\
\noalign{\smallskip}
\hline
$R/R_{\mathrm{Jup}}$ & 1.230 & 4   \\
$M/M_{\mathrm{Jup}}$ & 0.485 & 4     \\
$T_{\mathrm{14}}[\mathrm{h}]$ & 2.6064 & 1     \\  
\hline
    \hline
\end{tabular*}
(1)~\citet{2009A&A...501..785G};
(2) \citet{2019AJ....158..138S}; (3) \citet{2003tmc..book.....C,2006AJ....131.1163S}; (4) \citet{2015MNRAS.450.1760T}
\end{table}
This effect has made the PAH spectral signature a tracer of the star-formation rate \citep[e.g.][]{2016ApJ...818...60S}. PAHs are also thought to significantly impact the chemical and hydrodynamical evolution of protoplanetary discs \citep{2007prpl.conf..555D} and most likely atmospheric loss rates of close-in planets \citep{2009ApJ...705.1237G, 2020arXiv200508676M}. 

In addition to that, their standing in the astrobiology field grows because they might have played an important role as carbon sources for the synthesis of organic compounds, such as amino acids and nucleotides. \citet{Giese} experimentally demonstrated that no reaction of PAHs with ammonia and ammonia salts at temperatures of up to 150°C was detectable. However, it was observed before that PAHs can be pyrolyzed at elevated temperatures or even activated with Fenton’s reagent resulting in the formation of functionalized aromatic compounds \citep{TrappRef2,TrappRef1}. So far they have not been investigated in detail as potential photosensitizers in other photochemically driven reactions. Beyond using them as a carbon source they might have played a role in enabling transformations such as photoredox catalyzed alkylation of carbonyl compounds. PAHs can therefore be considered as both markers and carbon reservoirs for reduced and aromatic carbon compounds. 

In spite of their significance, until today, the technical capabilities of modern space and ground-based observatories have not been sufficient to detect these molecules in exoplanetary atmospheres. However, with ever-better telescopes such as the \textit{JWST} and its unmatched precision and accuracy, we may be closer than ever to finding them in these environments. \\ 

The production and destruction processes of PAHs are not well understood \citep{Kislov2013,2023A&A...678A..53D}. This makes it challenging to identify a sweet spot in the stellar-planetary parameter space to look for a promising candidate. However, a large number of studies have already been conducted by the \textit{Hubble Space Telescope} (\textit{HST}) and other facilities, which contain spectral information that could give a hint about the possible presence of PAHs. Most exoplanets recorded today are known to be either fully or partially cloudy or hazy \citep[e.g.][]{2023ApJS..269...31E,2023A&A...669A.142S,2023Natur.614..659R} which is usually expressed through a scattering component in the transmission and emission spectra for example a steep slope in the optical regime. But not all exoplanets show such a feature \citep[e.g.][]{2016Natur.529...59S}, and its origin is still under debate \citep[e.g.][]{2018AJ....156...38H,2018ApJ...865...80K,2020PSJ.....1...51H,2020NatAs...4..986H}. 

In our previous study \citep{2022MNRAS.512..430E}, we speculated that the origin of such slopes could be, at least partially, photochemically produced haze particles, such as PAHs. This speculation has been recently strengthened by \citet{2024NatAs...8..182H} who presented lab experiment-based evidence of aromatic compounds in photochemically produced hazes of exoplanets. Consequently, planets showing a strong slope component in their spectra might currently represent the best candidates to search for atmospheric PAHs. 

The target planet of this work, WASP-6~b, is a hot Saturn whose spectrum meets the criterion of a high slope (see e.g. Figures~\ref{Carter_BF1}~and~\ref{Carter_BF2}). The available dataset is a combination of observations from a recent analysis \citep[][hereafter referred to as C20]{2020MNRAS.494.5449C}. In this system, the host star is relatively quiet and the effect of stellar light contamination due to its heterogeneity on this slope would be minimal. Therefore, the feature is likely to have a planetary origin (C20). This gas giant, with around 50 per cent of the mass but 1.2 times the size of Jupiter, is orbiting its G8 type host star in about 3.4 days \citep{2009A&A...501..785G} making it a convenient target for observation. Besides, it is very similar to WASP-39 b which was already part of a successful ERS program in cycle 1 of the \textit{JWST} mission \citep[e.g.][]{2023Natur.614..659R}. 

Furthermore, it has an equilibrium temperature of around 1200~K (C20), a temperature regime at which molecular dynamics and thermochemical simulations suggest an onset of high concentration of thermalized PAHs \citep{2021CP....54811225H} and efficient PAH production from dehydrogenated benzene \citep{2020ApJ...900..188H}. Moreover in our most recent work \citep{2023A&A...678A..53D}, we used self-consistent 0-D simulations for a variety of PAH species to show that their abundances peak in these environments approximately at said temperatures. Lastly, the presence of an optical haze component in the atmosphere has already been proposed in C20 further indicating that WASP-6~b might be a favourable target for finding PAHs.\\ 

In this paper, we compare different sources for the optical slope and quantify their likelihood using the atmospheric retrieval method \citep{2018haex.bookE.104M} to show the possibility of a PAH presence and retrieve best-fitting models which are subsequently used as reference spectra. In the second part, we want to assess if \textit{JWST} provides the necessary precision to detect PAHs on WASP-6~b in the future. Accordingly, in section~\ref{Methods}, we lay out the retrieval assumptions and parameters as well as the process of creating the synthetic datasets. In section~\ref{Results}, we present and discuss our results divided into the analysis of the real and simulated observations. Finally, we conclude our work in section~\ref{Conclusion}.
\section{Methods}
\label{Methods}
In the following, we present the simulation setup. First, we describe the retrieval details, including the dataset, general assumptions made in petitRADTRANS, and the opacity structure. Finally, the production process of the synthetic observations is laid out.
\subsection{Observational data}
\label{obs}
We use the transmission spectrum from C20 as an input for our atmospheric retrievals. In the mentioned analysis, they corrected the data from observations of WASP-6 b for stellar heterogeneity effects. This was possible by modelling the stellar variability. Here, they reanalysed existing data from \textit{HST} STIS and \textit{Spitzer} IRAC which were already disclosed in \citet{2015MNRAS.447..463N}. The two \textit{Hubble} observations were conducted using G430L grism and G750L grism observing at 0.33-0.57 $\micron$ and 0.55-1.03 $\micron$, respectively. On the other side, \textit{Spitzer} was providing complementary data at 3.6 and 4.5 $\micron$. Additionally, they incorporated ground-based observations from the \textit{VLT} using FORS2 GRIS600B and GRISG600RI, both providing the bulk of the transit depths in the optical regime. Furthermore, the primary transit observation from an \textit{HST} G141 General Observer program was included capturing the spectral feature of water around 1.4 $\micron$. Finally, a single broadband filter image from the \textit{TESS} mission (0.6-1.0 $\micron$) completes the dataset \citep[for more details refer to][]{2020MNRAS.494.5449C}. 
\subsection{Atmospheric retrieval code petitRADTRANS}
\label{pRT}
In order to characterise exoplanetary atmospheres from transmission spectra, we conduct a set of atmospheric retrievals using the open source python package petitRADTRANS (pRT, version 3.0.4) \citep{2019A&A...627A..67M,2020A&A...640A.131M,2022A&A...665A.106A,2024JOSS....9.5875N}. For the forward modelling, it assumes a 1-D temperature and abundance structured atmosphere in hydrostatic equilibrium. pRT offers a variety of setups where one can choose different temperature profiles, equilibrium/disequilibrium or free chemistry, and a plethora of line and cloud opacities provided by commonly employed sources such as ExoMol \citep{2021A&A...646A..21C} and HITEMP \citep{2010JQSRT.111.2139R}. 

The code separately calculates the optical depth for grazing rays by summing up the contributions from each pressure layer up to the impact parameter \citep[for more information refer to][]{2019A&A...627A..67M}. Subsequently, the transmission functions are determined for each opacity species individually using the correlated-k approximation. As shown in Figure \ref{Opas}, the PAH spectral signature consists of a number of broad features i.e. this low-resolution treatment (c-k: $\lambda/\Delta\lambda$<1000) will be sufficient. Finally, from the product of the functions, the transmission spectrum is calculated.

In each retrieval run, the free parameter space of the respective model is explored using a Nested Sampling algorithm. Here, the code utilizes the Python wrapper PyMultiNest \citep{2014A&A...564A.125B} for MultiNest \citep{2008MNRAS.384..449F,2009MNRAS.398.1601F,2019OJAp....2E..10F}. This technique allows us to evaluate the quality of every model by comparing the evidences ln~$Z$ and reduced $\chi^2$.
\subsection{Main retrieval setup}
\label{Retrieval}
In summary, we use three main inputs for the forward modelling of the atmosphere: the pressure-temperature profile, basic planetary and stellar parameters, and the composition i.e. the opacity structure. First, we chose an isothermal/vertically constant temperature profile for which we define a uniform prior with limits of 300 and 1900~K. Furthermore, we assume planetary and stellar radii of 1.230 $R_{\mathrm{Jup}}$ and 0.87~$R_{\sun}$ respectively as well as a planetary mass of 0.485 $M_{\mathrm{Jup}}$ and therefore a surface gravity of log($g$) $\approx$ 2.90 (refer to Table~\ref{tab:W6b} for relevant values and references). These planet/star parameters are fixed in our retrievals since they are already determined with high precision from previous observations \citep{2009A&A...501..785G,2015MNRAS.450.1760T}. However, we do include a varying reference pressure P$_0$ defined as the height in the atmosphere where the local radius and gravity reach the fixed planetary values presented above. The respective prior is given by a log-uniform distribution with limits of $10^{-7}$ and 100 bar. The above parameters as well as the major infrared line species and continuum opacities described in the next section represent the foundation of our retrievals and remain unchanged for the whole analysis.
The essential free parameters for each run are given by the cloud/haze species with individual models including different types or combinations. An overview of all conducted runs is given in Table~\ref{tab:Setups}. In the current data analysis, we consider a total of 11 approaches. As mentioned, the abundances of the near-infrared line species as well as the temperature and reference pressure represent the basis of each free parameter set. These 8 quantities form the first Cloudless model. Moving on, we consider free parameters from a Gray Cloud, a Power Law Cloud, and a Haze Factor as well as their combinations as artificial cloud/haze opacities. Additionally, we study neutral PAHs (NPAHs), Tholins and their mixture. These parameters are described in section \ref{Haze/Clouds}. In the subsequent retrievals, following this analysis (HL, HC$_\mathrm{_{Setup}}$, HC, OR) no new models are introduced and the details are explained in the corresponding sections (ref. to section \ref{sec:PRISM}). Finally, 2000 live points and an evidence tolerance of 0.5 are used in all retrievals.
\begin{table}
    \centering
    \caption{Overview of the presented results in this work. CD: Current Data; C20: \protect\citet{2020MNRAS.494.5449C}; HL: Highest Likelihood; PRISM: \textit{JWST} NIRSpec PRISM; HC: Highest Consistent; HC$\mathrm{_{Setup}}$: Setup run for HC retrievals ; OR: Offset Retrievals. Refer to individual sections for more details.}
    \begin{tabular}{lccc}
        \hline
        \hline
        \multicolumn{4}{c}{Individual Retrieval Setups} \\
        \hline
        Name & Data  & Retrieved Models & Section\\
        \hline
        CD & C20 & 11&~\ref{sec: Carter} \\
        HL & C20 + PRISM & 3 &~\ref{Highest_Likelihood}\\
        HC$\mathrm{_{Setup}}$ & C20 & 1&\ref{Highest_Consistent} \\
        HC & C20 + PRISM & 3 &\ref{Highest_Consistent}\\
        OR & C20 &2& \ref{sec:Offset}\\
        \hline
        \hline
    \end{tabular}
    \label{tab:Setups}
\end{table}
\subsection{Line species/continuum opacities}
\label{Lines}
The core of the opacity structure is given by major infrared absorbers constituted of line species and continuum opacities. In pRT, the molecular abundances are implemented as relative mass fractions. Since there is already strong evidence for water, sodium, and potassium in the atmosphere of WASP-6~b (refer to C20) we included these in our retrievals. Additionally, we account for CO, CO$_2$ and CH$_4$ because of their prominent features in the near-infrared and common presence in these environments. For H$_2$O, CO, CO$_2$ and CH$_4$ we chose the opacity calculations from ExoMol \citep[][respectively]{2018MNRAS.480.2597P,2015ApJS..216...15L,2020MNRAS.496.5282Y,2024MNRAS.528.3719Y}, and Na and K data from the Vienna Atomic Line Database \citep{1995A&AS..112..525P} using line broadening from \citet{1996MNRAS.283..821S} \citep[for details see][]{2019A&A...627A..67M}. Furthermore, we consider the Rayleigh scattering species H$_2$ and He \citep[][respectively]{1962ApJ...136..690D,1965PPS....85..227C} as well as collision-induced absorption (CIA) contributions arising from H$_2$-H$_2$, and H$_2$-He interactions \citep[][and references therein]{1988ApJ...326..509B,1989ApJ...336..495B,1989ApJ...341..549B,2001JQSRT..68..235B,2002A&A...390..779B,2012JQSRT.113.1276R}. For the line species, we construct vertically constant, log-uniform priors between $10^{-21}$ and 0.1. Finally, the remaining mass fraction is split between 76.6 per cent H$_2$ and 23.4 per cent He (solar).
\subsection{Haze/cloud opacities}
\label{Haze/Clouds}
For this work, the clouds/hazes are the essential opacity sources since we aim to compare their impact on the transmission spectrum and characterize the optical slope of WASP-6~b in order to evaluate if PAHs will be detectable in future observations. In addition to PAHs, a few different synthetic clouds described in \citet{2019A&A...627A..67M} and their combinations are considered. Furthermore, we implement another molecule group similar to PAHs namely Tholins. They are described further in section~\ref{PT}.
\subsubsection{Gray Clouds}
\label{Gray}
This simple cloud type is often resorted to in atmospheric retrievals \citep[see e.g.][]{2015ApJ...814...66K} and is used to mimic highly opaque material. Here, we define a certain height in the atmosphere i.e. a pressure layer where the gray cloud top is located at. Below this sheet, the atmosphere becomes completely opaque at all wavelengths implemented using unnaturally high opacities for all deeper layers. Similar to the reference pressure, we construct a log-uniform prior with limits of $10^{-7}$ and 100 bar.
\subsubsection{Power Law Clouds}
\label{PLC}
Power Law Clouds introduce a slope opacity in the transmission spectrum with variable strength and steepness. They are defined by the following opacity function
\begin{equation}
    \kappa \ = \ \kappa_\mathrm{0} \ \times\ \bigg(\frac{\lambda}{\lambda_\mathrm{0}}\bigg)^{\gamma_{\mathrm{scat}}}
\end{equation}
where $\kappa_\mathrm{0}$ and $\gamma_{\mathrm{scat}}$ are free parameters and $\lambda_\mathrm{0}$ is fixed at 0.35 $\micron$. This treatment allows us to account for simple, linear slopes of any origin. The priors are given by log-uniform and uniform distributions between $10^{-8}$ and $10^{4}$ cm$^2$/g as well as $-20$ and $0$ for $\kappa_\mathrm{0}$ and $\gamma_{\mathrm{scat}}$, respectively.
\subsubsection{Haze Factor}
\label{Haze}
A third synthetic cloud/haze option is the Haze Factor that scales up the Rayleigh scattering contribution according to 
\begin{equation}
    \kappa \ = \ f_{\mathrm{Haze}}\ \times \ \kappa_{\mathrm{Rayleigh}}
\end{equation}
similar to scattering hazes observed on Earth. We construct uniform priors between $1$ and $60$ times Rayleigh scattering strength. 
\subsubsection{PAH and Tholin opacities}
\label{PT} 
In the following, we compare the synthetic slope sources described above with the impact of 'real' molecules/particles like PAHs and Tholins. Tholins define another group of large, complex hydrocarbons and are as such similar to PAHs \citep[e.g.][]{1979Natur.277..102S}. Since the production mechanisms of both species are likely of photochemical origin they might coexist in exoplanetary atmospheres just like on Saturn's moon Titan. Thus, we additionally consider Tholins in our retrievals. \\

As already mentioned, PAHs are primarily identifiable by their distinct absorption features in the infrared spectrum. The PAH absorption cross-sections were calculated as in \citet{2001ApJ...554..778L} \citep[updates from][]{2007ApJ...657..810D} following our previous study \citep[][more details in \citet{2016A&A...586A.103W}]{2022MNRAS.512..430E} for neutral circumcoronene. This molecule contains 54 carbon and 18 hydrogen atoms bound exclusively in the form of benzene rings with an overall hexagonal shape. Figure 1 in \citet{2022MNRAS.512..430E} illustrates the subsequent cross-sections for neutral and ionized PAHs. These cross-sections, commonly employed in protoplanetary disc characterizations, reflect optical properties of 'astroPAHs' and are in agreement with results from laboratory experiments and ISM observations.

Both neutral and charged cross-sections show distinct characteristics with prominent features at 3.3 $\micron$ and beyond 5 $\micron$ and within the reach of MIRI LRS. These features, and especially the ones at 6.2 $\micron$ and 7.8 $\micron$, have been targeted with previous disc surveys by \textit{Spitzer} and \textit{Herschel} telescopes. However, charged PAHs are expected to contribute to the mid-infrared peaks more and are not likely to be present on exoplanets at the level of neutral PAHs. Moreover, in discs, silicate features would contaminate them, making their detection more difficult \citep{2017ApJ...835..291S,2022MNRAS.512..430E}. Nevertheless, this might not be the case for hot exoplanets. Yet, in principle, only the optical slope short-ward of 1 $\micron$ is significantly covered in the transmission spectrum of WASP-6~b and thus represents the only trademark we could identify using the current dataset. 

Finally, on Earth and in discs, PAHs are known to form particle clusters of sizes up to a few $\micron$ which is why we also include a scattering component based on classical Mie theory as part of the PAH opacity. Here, we use the standard Mie formalism for calculating the scattering cross-sections \citep[as in][]{2022MNRAS.512..430E}. As mentioned, Tholins are similar in their composition to PAHs. However, they are in general not constituted of benzene rings which create the distinct set of PAH features in the near and mid-infrared. Thus, we resort to a full Mie theory treatment for both the absorption and scattering cross-sections for Tholins as in \citet{2020A&A...642A.173N}. \\
\begin{figure}
	\includegraphics[width=\columnwidth]{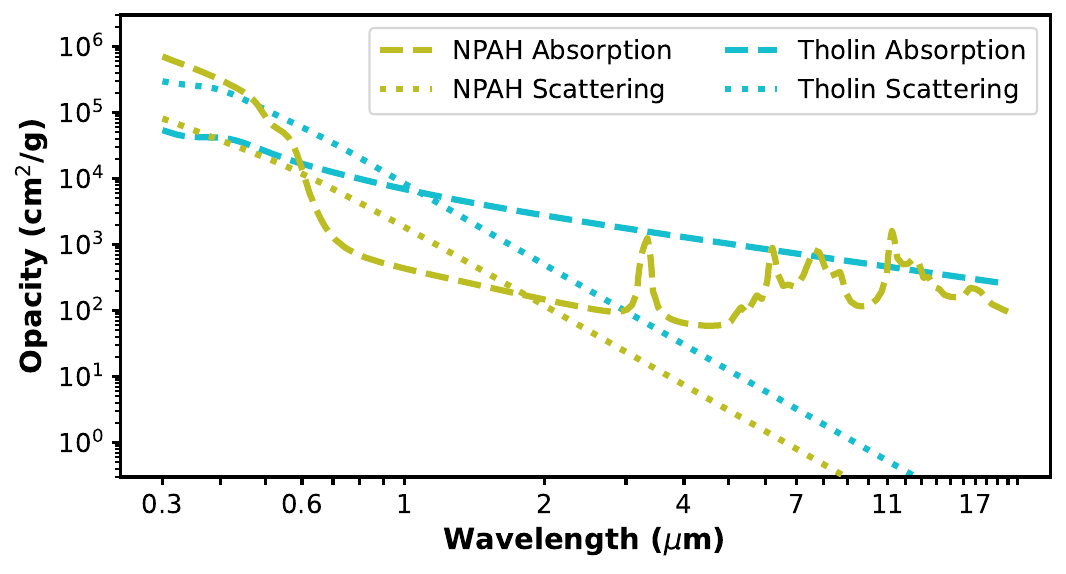}
    \caption{Absorption (dashed) and scattering (dotted) opacities in cm$^2$/g for both neutral PAHs (yellow) and Tholins (cyan). Here, a log-normal distribution with a $\sigma_{\mathrm{log-normal}}$ of 1.05 and a mean radius of 0.1 $\micron$ is used.}
    \label{Opas}
\end{figure}

In conclusion, for neutral PAHs, we include monodisperse absorption from circumcoronene and size-dependent scattering to account for the \citet{2001ApJ...554..778L} cross-sections mentioned above as well as the contribution from particle accumulations. This treatment is identical to the implementation in \citet{2022MNRAS.512..430E}. Tholins on the other hand are included in a completely size-dependent manner. In our final retrieval runs, we fix the particle sizes for both groups to a log-normal distribution with a $\sigma_{\mathrm{log-normal}}$ of 1.05 and a mean radius of 0.1 $\micron$. The resulting absorption and scattering opacities are shown in Figure~\ref{Opas} for both molecule groups. Furthermore, we consider their contribution to the mean molecular weight of the atmosphere and retrieve their abundances as in the case of regular line species using log-uniform and vertically constant priors between $10^{-21}$ and 0.1. For Tholins, we assume light hydrocarbons as in Titans upper atmosphere with a mass of 330 amu which is consistent with observed particle sizes.
\begin{figure*}
    \begin{subfigure}{0.49\textwidth}
        \centering
        \begin{tikzpicture}
            \node[anchor=south west,inner sep=0] (image) at (0,0) {\includegraphics[width=\textwidth]{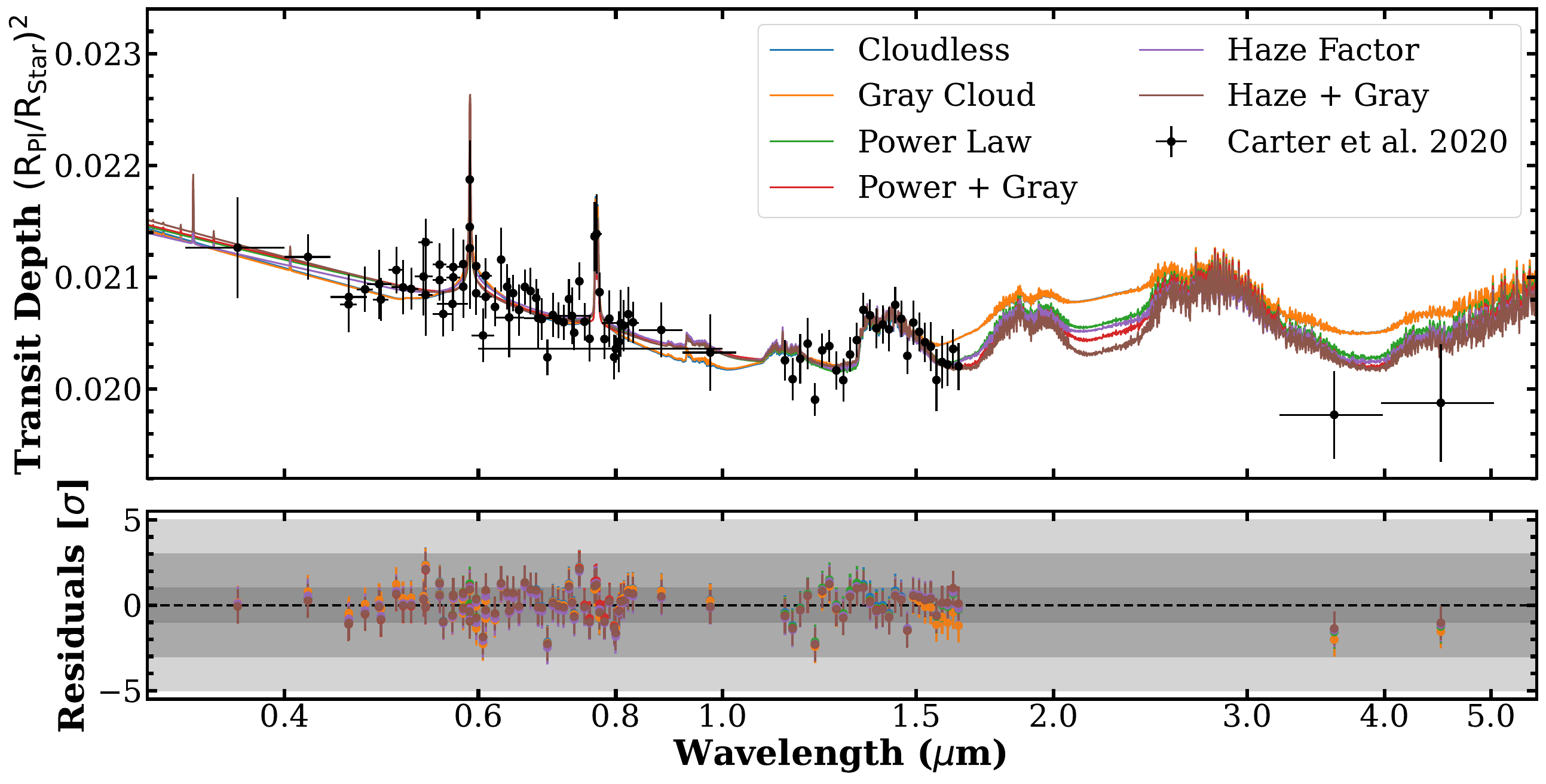}};
            \begin{scope}[x={(image.south east)},y={(image.north west)}]
            \node [anchor=north west] at (0.12,0.97) {(a)};
            \end{scope}
        \end{tikzpicture}
        \phantomcaption
        \label{Carter_BF1}
    \end{subfigure}
    \begin{subfigure}{0.49\textwidth}
        \centering
	    \begin{tikzpicture}
            \node[anchor=south west,inner sep=0] (image) at (0,0) {\includegraphics[width=\textwidth]{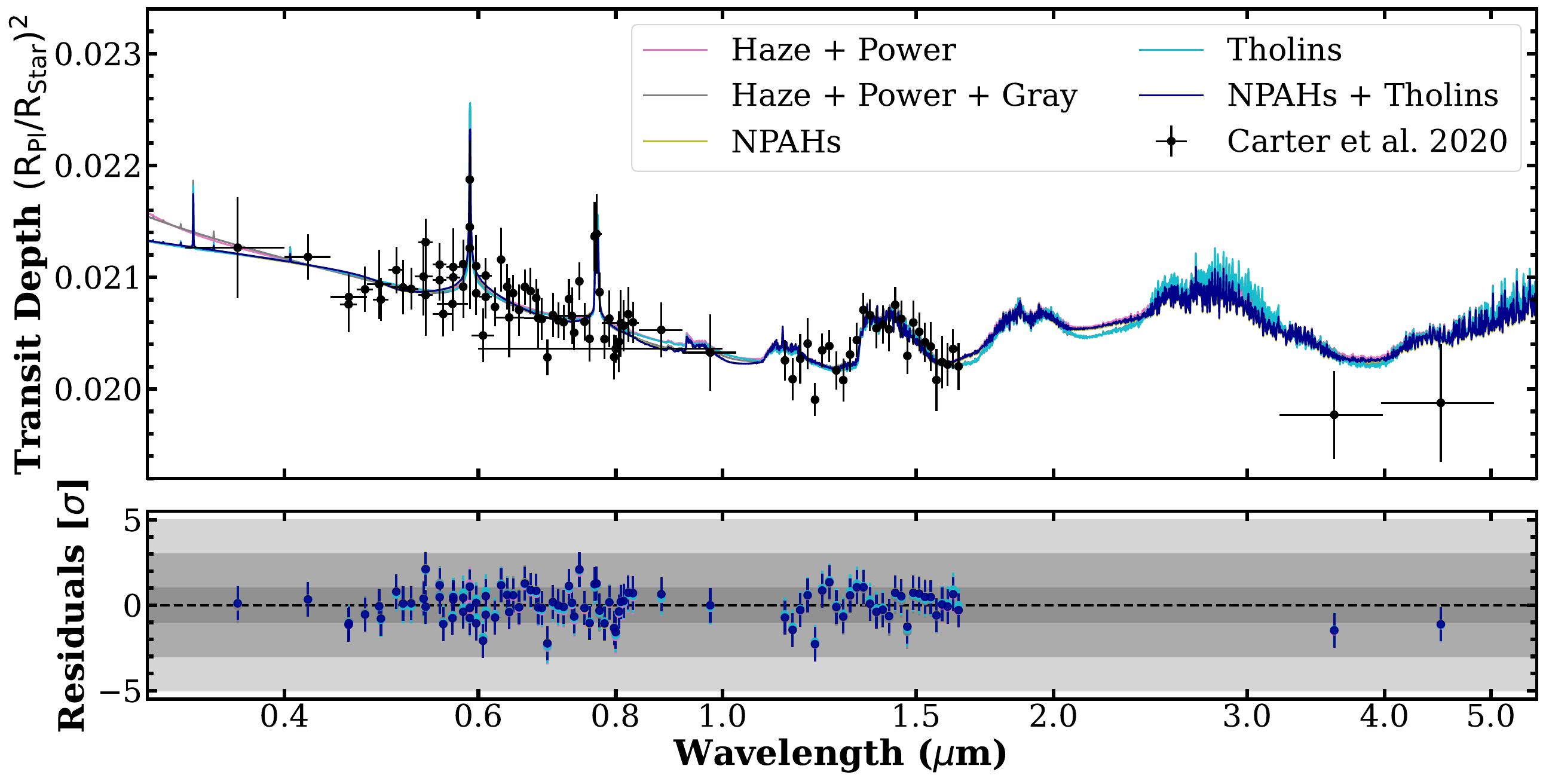}};
            \begin{scope}[x={(image.south east)},y={(image.north west)}]
            \node [anchor=north west] at (0.12,0.97) {(b)};
            \end{scope}
        \end{tikzpicture}
        \phantomcaption
        \label{Carter_BF2}
    \end{subfigure} 
   \begin{subfigure}{0.49\textwidth}
        \centering
        \begin{tikzpicture}
            \node[anchor=south west,inner sep=0] (image) at (0,0) {\includegraphics[width=1\textwidth]{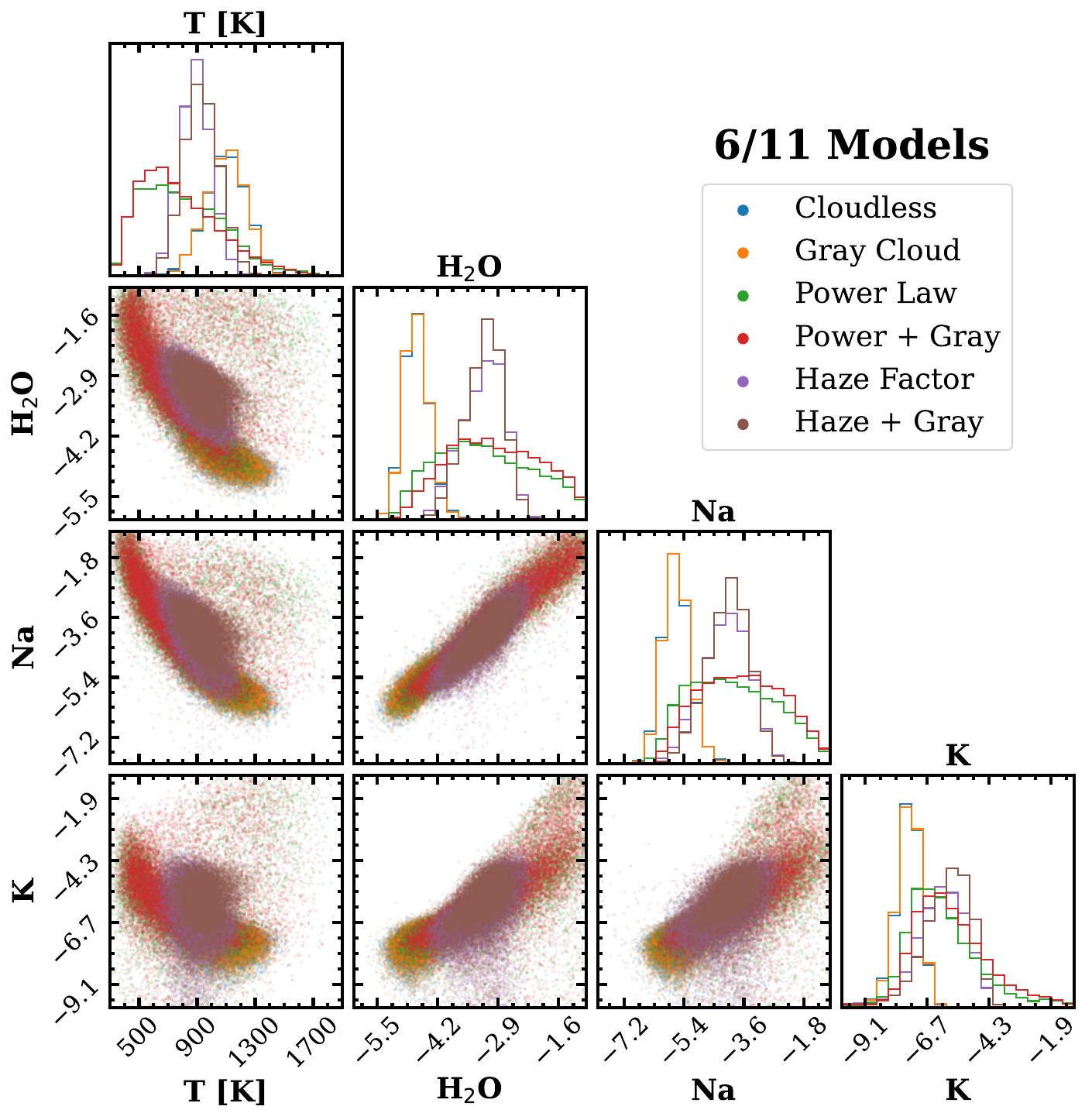}};
            \begin{scope}[x={(image.south east)},y={(image.north west)}]
            \node [anchor=north west] at (0,1) {(c)};
            \end{scope}
        \end{tikzpicture}
        \phantomcaption
        \label{Carter_Post1}
    \end{subfigure}
    \begin{subfigure}{0.49\textwidth}
        \centering
        \begin{tikzpicture}
            \node[anchor=south west,inner sep=0] (image) at (0,0) {\includegraphics[width=1\textwidth]{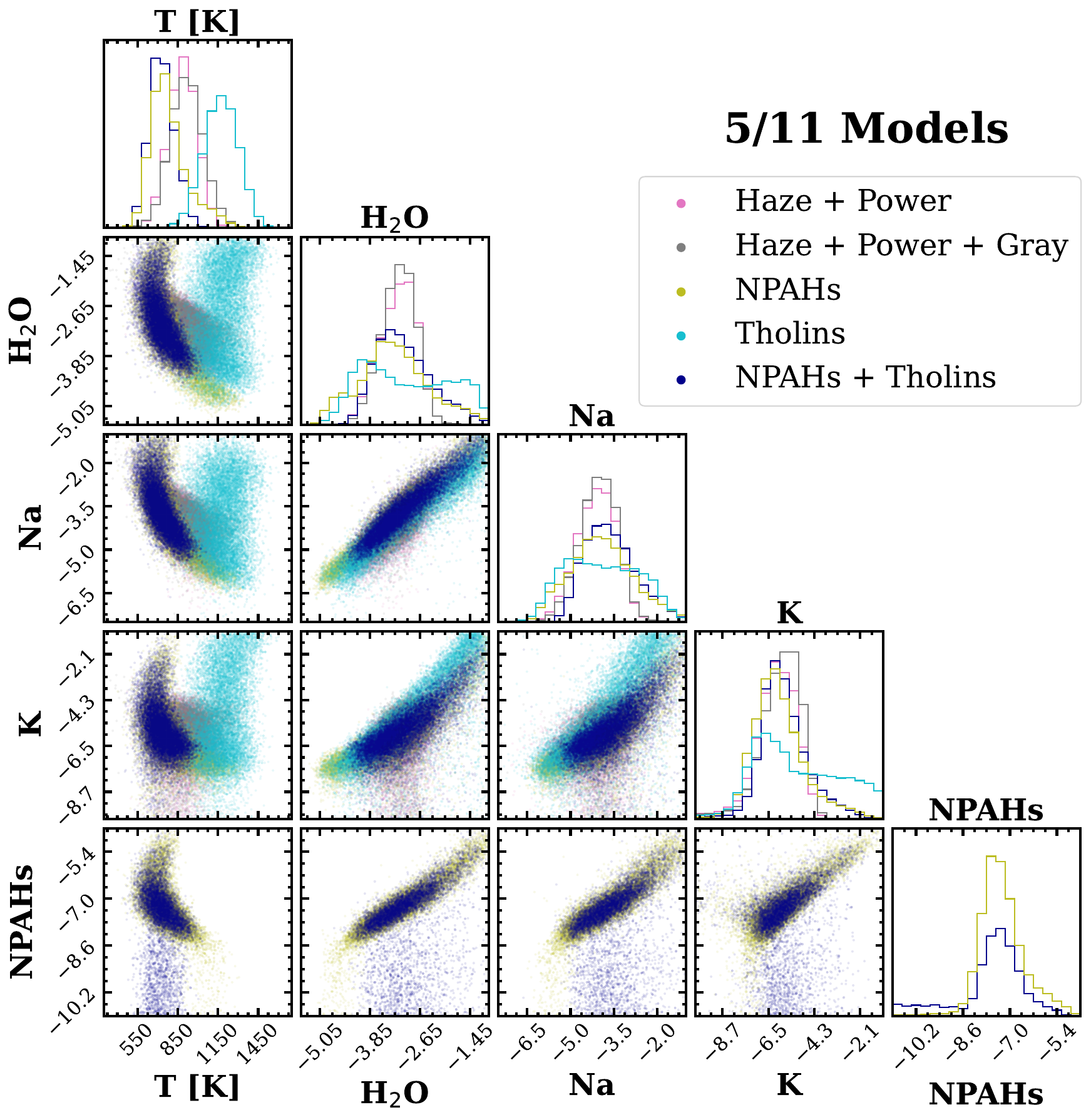}};
            \begin{scope}[x={(image.south east)},y={(image.north west)}]
            \node [anchor=north west] at (0,1) {(d)};
            \end{scope}
        \end{tikzpicture}
        \phantomcaption
        \label{Carter_Post2}
    \end{subfigure}  
    \caption{Results for the CD (refer to Table \ref{tab:Setups}) retrieval runs. Subfigures (a) and (b): Top panels: Best-fitting spectra for the 11 models including the currently available data (C20) in black. Bottom panel: Residuals are plotted for each model in the respective colors showing the deviation of the data points from the spectra in units of the corresponding transit depth error ($\sigma$). The shaded bars indicate the 1, 3 and 5 $\sigma$ regions. Subfigures (c) and (d): Corner plots including the temperature, H$_2$O, Na, K and PAH abundances. The 1D posteriors are presented as histograms on the diagonal and the 2D distributions are displayed by sample points.}
    \label{CarterRes}
\end{figure*}
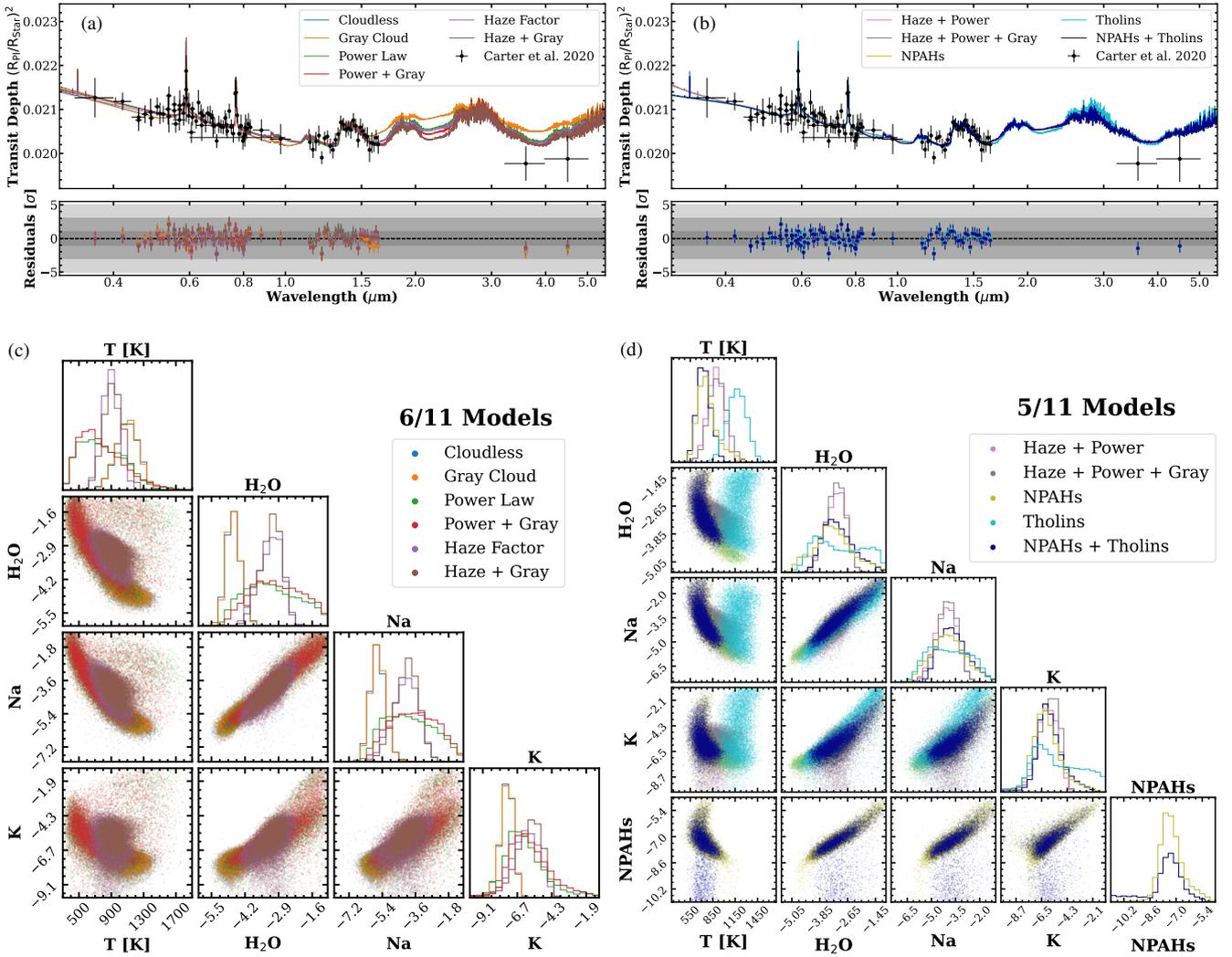
\subsection{Simulated observations}
\label{SimObs}
In the second part of this work, we create synthetic datasets for the \textit{JWST} NIRSpec mode PRISM considering a single transit observation. The results from the PAH retrievals on the current data are used as input for \textit{JWST}'s Exposure Time Calculator (ETC) (for more details ref. to section \ref{sec:PRISM}). Here, we utilize the official ETC PandExo 3.0 \citep{2017PASP..129f4501B} to create datasets including uncertainties for this instrument mode. Table~\ref{tab:W6b} shows the planetary and stellar \citep[model from PHOENIX:][]{2013A&A...553A...6H} parameters necessary for the calculation involving quantities like the transit duration and the stellar magnitude. Additionally, another 2 hours before and after the transit are assumed for the baseline. We choose NIRSpec PRISM (hereafter also referred to as PRISM) since it covers most of the PAH spectral features in the near-infrared including part of the slope as well as the 3.3 $\micron$ peak \citep[refer to Figure 1 in][]{2022MNRAS.512..430E} and provides a resolution of R $\sim$ 100 that would be sufficient to capture the broad PAH features.
\section{Results and Discussion}
\label{Results}
In the following, we will present the retrieval results on the currently available data as well as simulated datasets for NIRSpec PRISM (refer to Table~\ref{tab:Setups}).
\subsection{Retrievals on currently available data}
\label{sec: Carter}
\begin{figure*}
	\includegraphics[width=\textwidth]{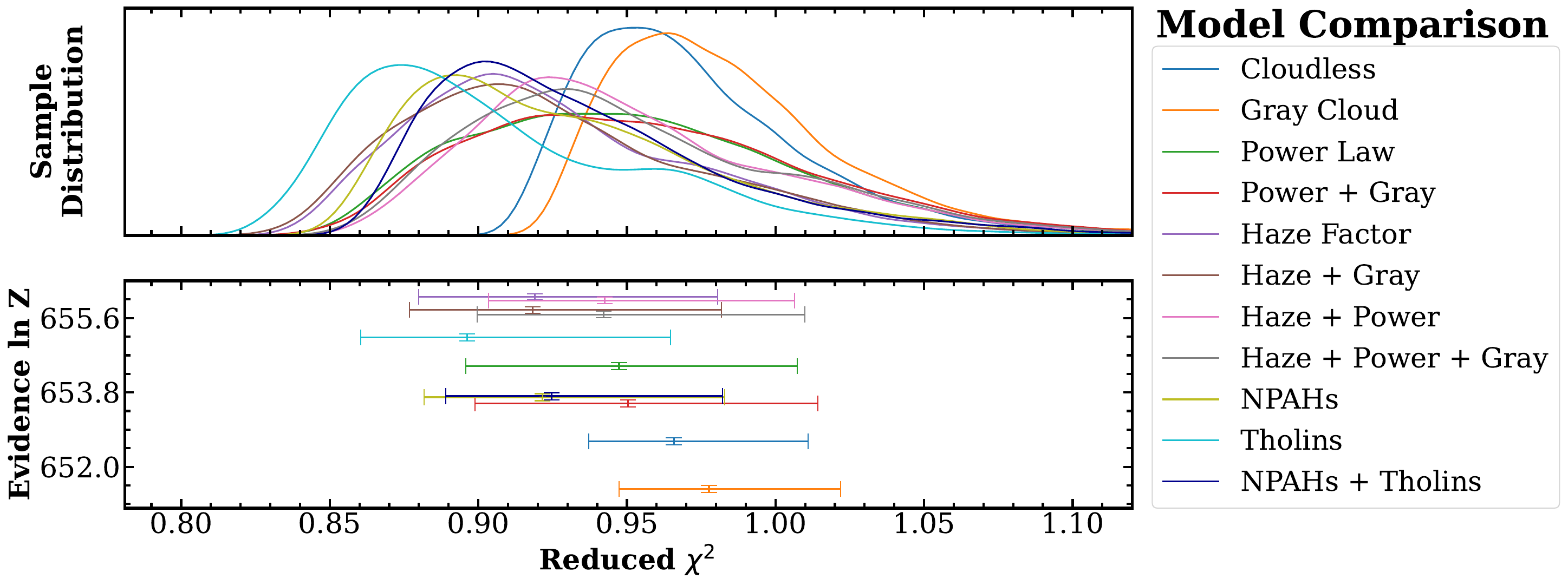}
    \caption{Top panel: Reduced $\chi^2$ sample distributions for the CD retrievals. Bottom panel: Reduced $\chi^2$ plotted against the evidence. The uncertainties on the $\chi^2$ are given by the 16 and 84 per cent quantiles of the distributions while the errors on the evidences are directly produced by MultiNest.}
    \label{Carter_Chi}
\end{figure*}

\begin{figure}
    \includegraphics[width=\columnwidth]{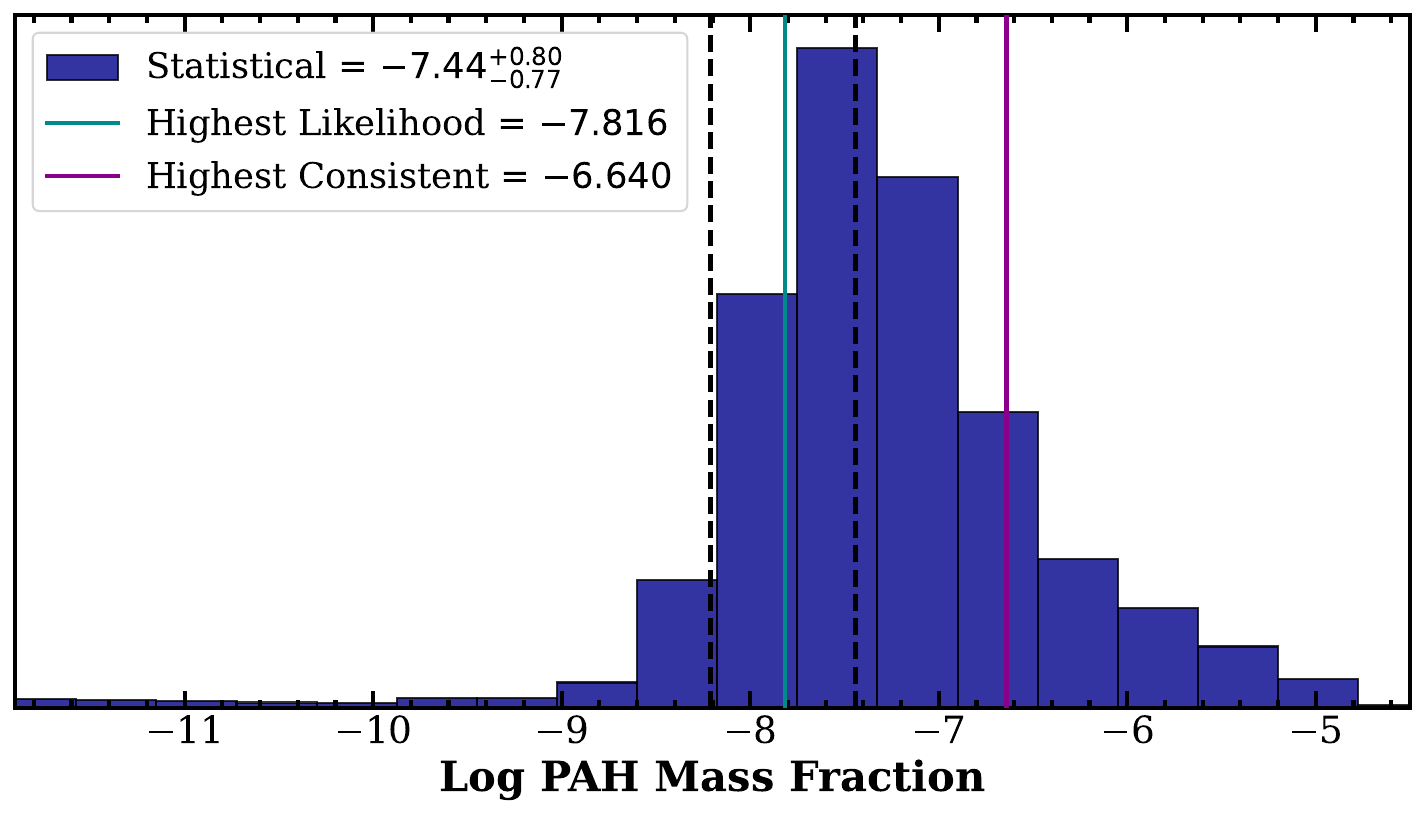}
    \caption{Posterior distribution of the PAH mass fraction from the CD retrieval. The 16 and 84 per cent quantiles as well as the medians are shown as dashed lines. Finally, the best-fitting PAH abundance from the original retrieval as well as the value used in the second part of the PRISM analysis (refer to section \ref{sec:PRISM}) are shown in cyan and magenta, respectively.}
    \label{PAH_MFs}
\end{figure}

The results are displayed in Figures~\ref{CarterRes}~to~\ref{PAH_MFs}. In the top panels of Figures~\ref{Carter_BF1} and~\ref{Carter_BF2}, the best-fitting spectra i.e. the spectra with the highest likelihoods are presented for each of the 11 models. The currently available data including the prominent optical slope are shown in black. Additionally, the model spectra are binned to the data and the resulting deviations are computed. These residuals are shown in units of the corresponding transit depth uncertainties ($\sigma$) in the bottom panels. According to the Figures, as expected, the models overlap predominantly in the optical regime and around the water feature at 1.4 $\micron$ i.e. in the region covered by the data. However, we observe large differences between the spectra at longer wavelengths starting around 1.5 $\micron$. This already indicates that complementary data are essential to be able to distinguish between the different models. Conveniently, NIRSpec PRISM would be able to fill the data gap thanks to its wavelength coverage. 
\\ \\
In addition, for most spectra, we observe a rise in transit depth around 2-3 $\micron$ caused by the CIA species. This effect is so prominent because of the low abundance of line species found in the atmosphere, leading to a dominating presence of hydrogen and helium (refer to Figure \ref{Full_Carter_Post} as well as Tables~\ref{appendix:tab1} and~\ref{appendix:tab2} for details). The Cloudless and Gray Cloud approaches show the largest peaks i.e. a high H$_2$ and He content. Presumably, this is a result of the optical slope which needs to be modelled by strong Rayleigh scattering, assuming the absence of any other slope sources. Since this effect is not observed to the same extent on WASP-39 b \citep[see e.g.][]{2023Natur.614..659R}, which has similar properties, the true shape of the transmission spectrum might differ significantly from all model spectra. Nonetheless, in this analysis, we will proceed with these assumed compositions.
In Figures~\ref{Carter_Post1} and~\ref{Carter_Post2}, we show the posterior distributions for the retrieved temperature and the abundances of the three robustly detected molecules H$_2$O, Na, and K as well as PAHs in a corner plot. The full 1D posteriors are given in Figures \ref{Full_Carter_Post} and \ref{Full_Carter_Cloud_Post} in the appendix. Accordingly, the medians of the temperature distributions range from approximately 700 to 1200 Kelvin with a combined mean of around 900 Kelvin. Here, the models are divided into three groups with approaches including Power Law Clouds or PAHs as the dominating cloud opacity preferring lower temperatures. On the other side, medium temperatures are adopted by Haze Factors and high temperatures by Tholins, and slope-free models (Cloudless and Gray Cloud).

Furthermore, we observe that the inclusion of slope sources leads to increased mass fractions for the line species. As already mentioned, this is presumably attributed to the need for high Rayleigh scattering i.e. comparatively low values in slope-free cases. However, another contributing factor could be the 'flattening effect' of the additionally introduced cloud/haze opacities which could overpower molecular features. To maintain the same feature shape, the retrieval likely responds to this effect in general with higher abundances. In some cases, this causes a degeneracy which becomes apparent in the corner plots. The posteriors of the Power Law Clouds, PAHs and Tholins show broad and in some cases almost plateau-like distributions for the line species which is not observed to the same extent for the Haze Factor nor the slope-free approaches. 

The 2D plots reveal a correlation between higher mass fractions and higher PAH abundances. Additionally, the Tholin model (similar for PAHs) exhibits a near independence between the temperature and the mass fractions within the plateau-like regions. Thus, assuming a constant temperature, linearly increasing the mass fractions (log-log space) of both line and counteracting cloud species leads to spectra with similar likelihoods presumably due to the ratio of the feature strengths between the molecules remaining approximately constant. On the other side, the same might not be the case for the Haze Factor and slope-free models as the impact of Rayleigh scattering is limited to the lower wavelength regime whereas PAHs, Tholins and the Power Law Cloud exhibit a more 'broad-chromatic' influence. 

As we will demonstrate in the next section, complementary data allow us to lift this degeneracy. However, before we proceed, we want to highlight a few additional insights coming from this initial analysis. Firstly, the PAHs plus Tholins model shows a more shallow peak in the PAH mass fraction. The retrieval finds a solution where both cloud types are included showing that a coexistence of PAHs and Tholins could be possible (see Figure \ref{Full_Carter_Cloud_Post}). Furthermore, we do not identify any mentionable signal of CO, CO$_2$ and CH$_4$ in the transmission spectrum and the reference pressures (P$_0$) are roughly ranging from 0.01 to 1 bar. For details on these quantities refer to Tables~\ref{appendix:tab1} and~\ref{appendix:tab2} as well as Figures \ref{Full_Carter_Post} to \ref{Full_Carter_Cloud_Post}. 

Additionally, each model quality is summarized by the reduced $\chi^2$-evidence plot in the lower panel of Figure~\ref{Carter_Chi}. Here, we derived the reduced $\chi^2$ uncertainties by computing the 16 and 84 per cent quantiles along with the median from the sample distributions shown in the top panel of the Figure. The errors on the evidences ln~$Z$ are generated by the MultiNest algorithm. It is important to note that the reduced $\chi^2$ comparison should be approached with caution primarily due to the challenges associated with accurately estimating the degrees of freedoms in multidimensional parameter spaces \citep[see e.g.][]{2010arXiv1012.3754A}. Nonetheless, we use this quantity as an additional metric to assess the quality of the different models. This approach ensures that we do not solely rely on the evidence which can also be influenced by the properties of said parameter space for example the choice of the prior distributions. To maximize the compatibility of these values, we chose not to compare the best-fitting reduced $\chi^2$ but, as mentioned, compute the median and 1$\sigma$ distributions from the sampled posteriors. Additionally, the number of points significantly outweighs the number of free parameters in our retrievals which further increases the reliability of the results. 

Therefore, when considering both these metrics, we observe that all of the cloud/haze opacities, except for the Gray Cloud, show a higher quality (higher evidence, lower $\chi^2$) in comparison to the Cloudless approach. While most of the $\chi^2$ medians agree with each other's 1$\sigma$ intervals the evidences are clearly not with the best models including the Haze Factor.
Interestingly, Tholins, with the lowest $\chi^2$ and higher evidence than PAHs, stand out once again. Considering that the cross-sections are completely based on Mie theory this motivates a more comprehensive scattering analysis. Nevertheless, the evidences of all models are still comparable with a maximum difference between the cloud/haze scenarios of around $\Delta \mathrm{lnZ = 4}$. Overall, both quality metrics show a slight preference of the cloud models over the slope-less approaches. However, no model is strongly favoured, instead all test cases seem similarly plausible.

Finally, in Figure~\ref{PAH_MFs}, the PAH posterior distribution, including median and 16/84 per cent quantiles as well as the highest likelihood value is once more presented. Therefore, in the best-fitting scenario for the currently available data a relative mass abundance of around $10^{-7.8}$ is expected to be present in the atmosphere.

We conclude, that PAHs are a possible source for the optical slope in the transmission spectrum and thus might be present in WASP-6~b's atmosphere. However, according to both quality metrics, other models are also able to replicate the data equally well. Hence, an unambiguous PAH detection is not possible on WASP-6~b with the current data including only the optical slope component. 

We now approach the question of whether additional data covering a wider wavelength range could provide constraints on the PAH abundance on WASP-6~b. To this aim, we assess whether \textit{James Webb's} near-infrared spectrograph NIRSpec would be capable of facilitating a clear detection of PAHs assuming they are the dominant haze species in the atmosphere of the planet. 
\subsection{Detectability with NIRSpec PRISM}
\label{sec:PRISM}
\begin{figure*}
    \begin{subfigure}{0.49\textwidth}
        \centering
        \begin{tikzpicture}
            \node[anchor=south west,inner sep=0] (image) at (0,0) {\includegraphics[width=\textwidth]{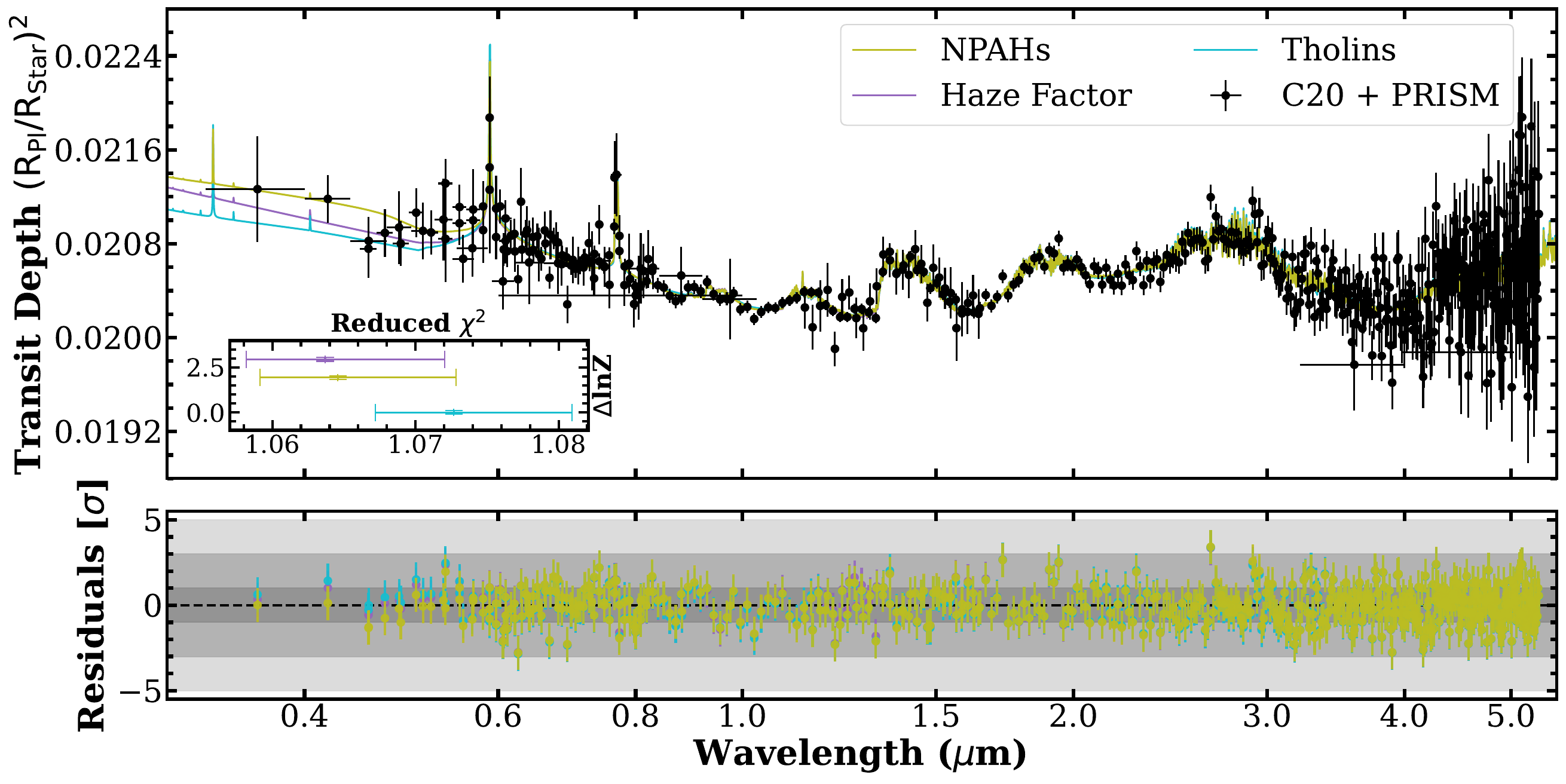}};
            \begin{scope}[x={(image.south east)},y={(image.north west)}]
            \node [anchor=north west] at (0.12,0.97) {(a)};
            \end{scope}
        \end{tikzpicture}
        \phantomcaption
        \label{fig:CP}
    \end{subfigure}
    \begin{subfigure}{0.49\textwidth}
        \centering
	    \begin{tikzpicture}
            \node[anchor=south west,inner sep=0] (image) at (0,0) {\includegraphics[width=\textwidth]{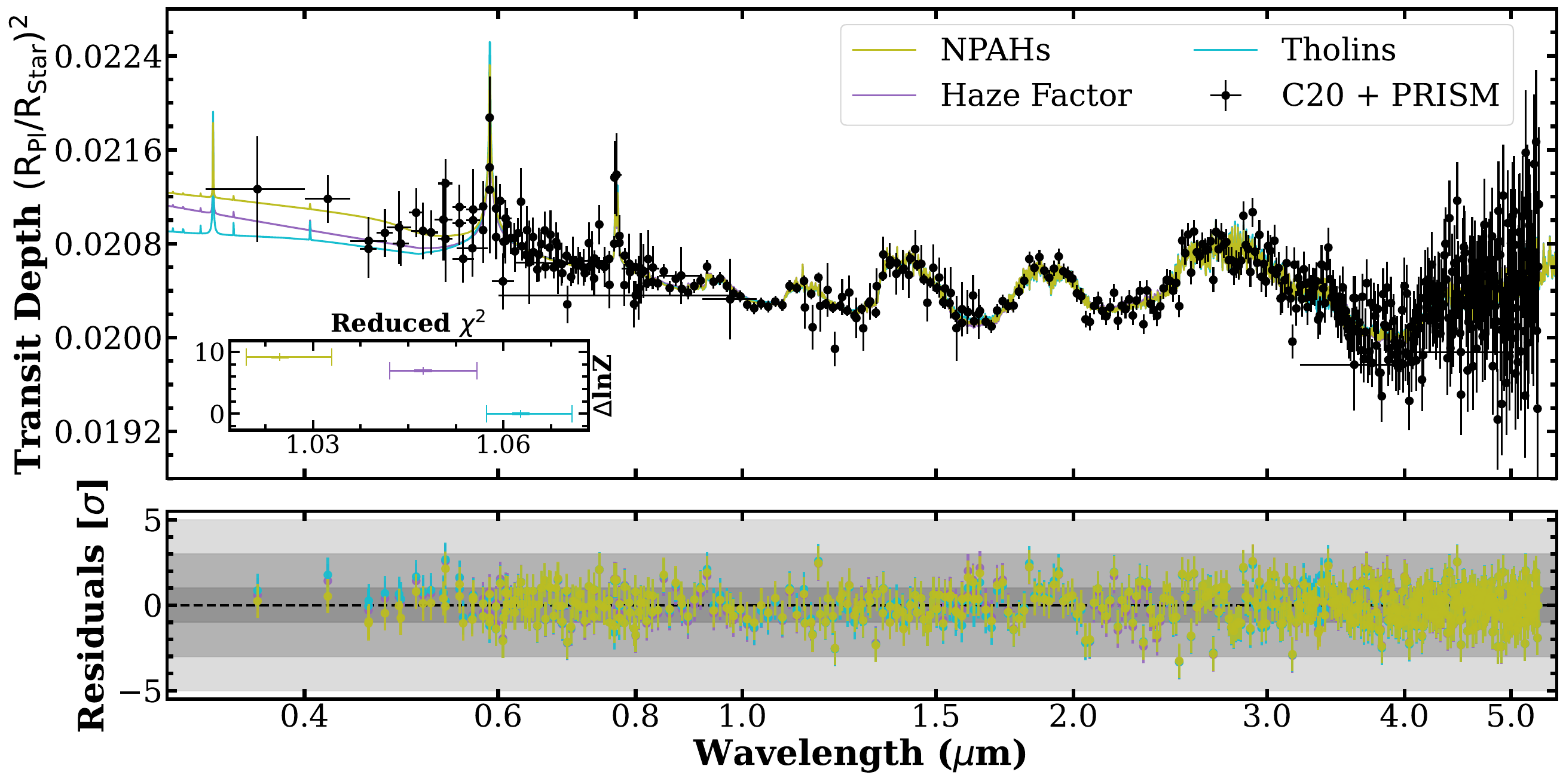}};
            \begin{scope}[x={(image.south east)},y={(image.north west)}]
            \node [anchor=north west] at (0.12,0.97) {(b)};
            \end{scope}
        \end{tikzpicture}
        \phantomcaption
        \label{fig:CPN}
    \end{subfigure} 
   \begin{subfigure}{0.49\textwidth}
        \centering
        \begin{tikzpicture}
            \node[anchor=south west,inner sep=0] (image) at (0,0) {\includegraphics[width=1\textwidth]{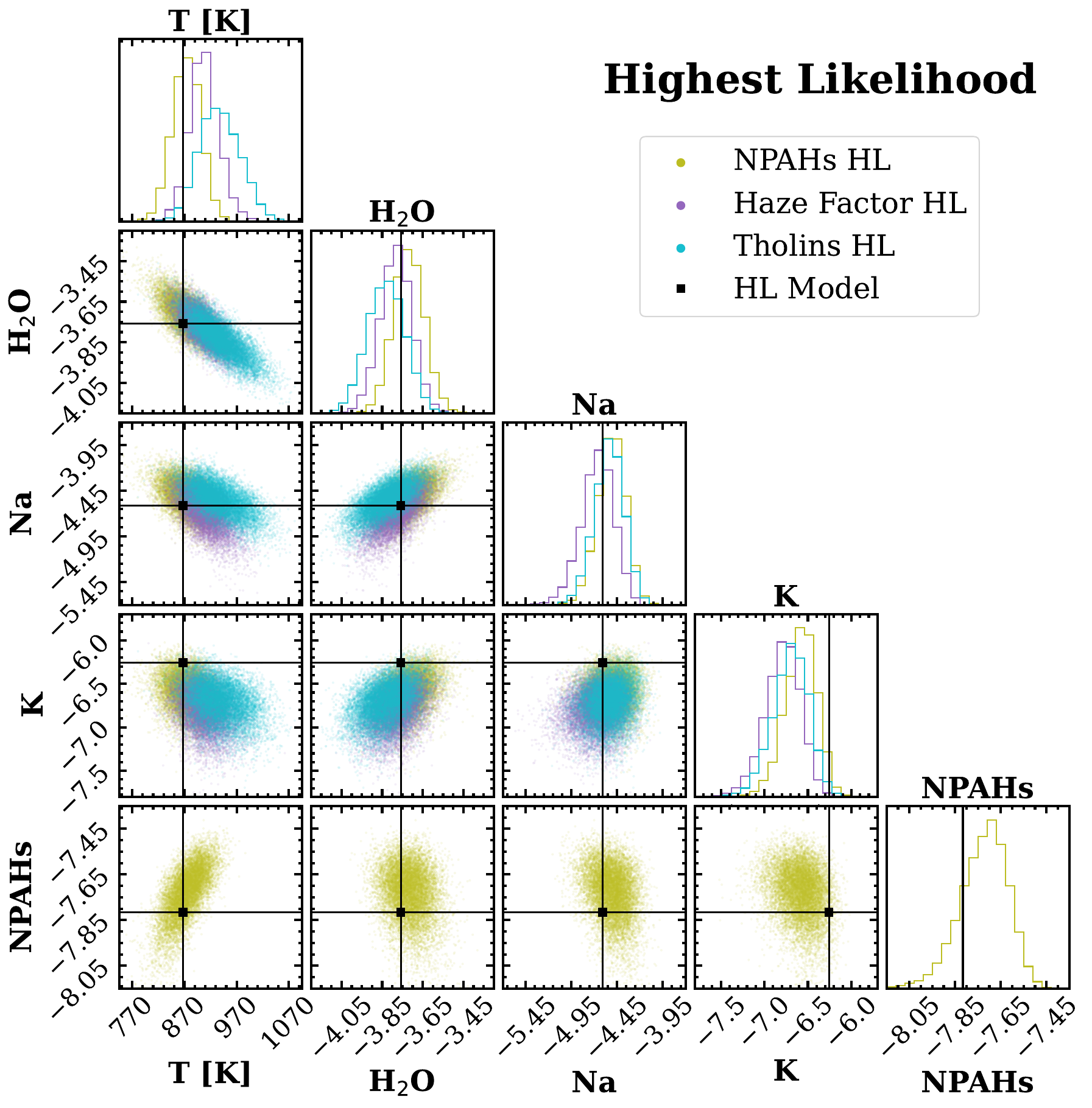}};
            \begin{scope}[x={(image.south east)},y={(image.north west)}]
            \node [anchor=north west] at (0,1) {(c)};
            \end{scope}
        \end{tikzpicture}
        \phantomcaption
        \label{fig:HighLike}
    \end{subfigure}
    \begin{subfigure}{0.49\textwidth}
        \centering
        \begin{tikzpicture}
            \node[anchor=south west,inner sep=0] (image) at (0,0) {\includegraphics[width=1\textwidth]{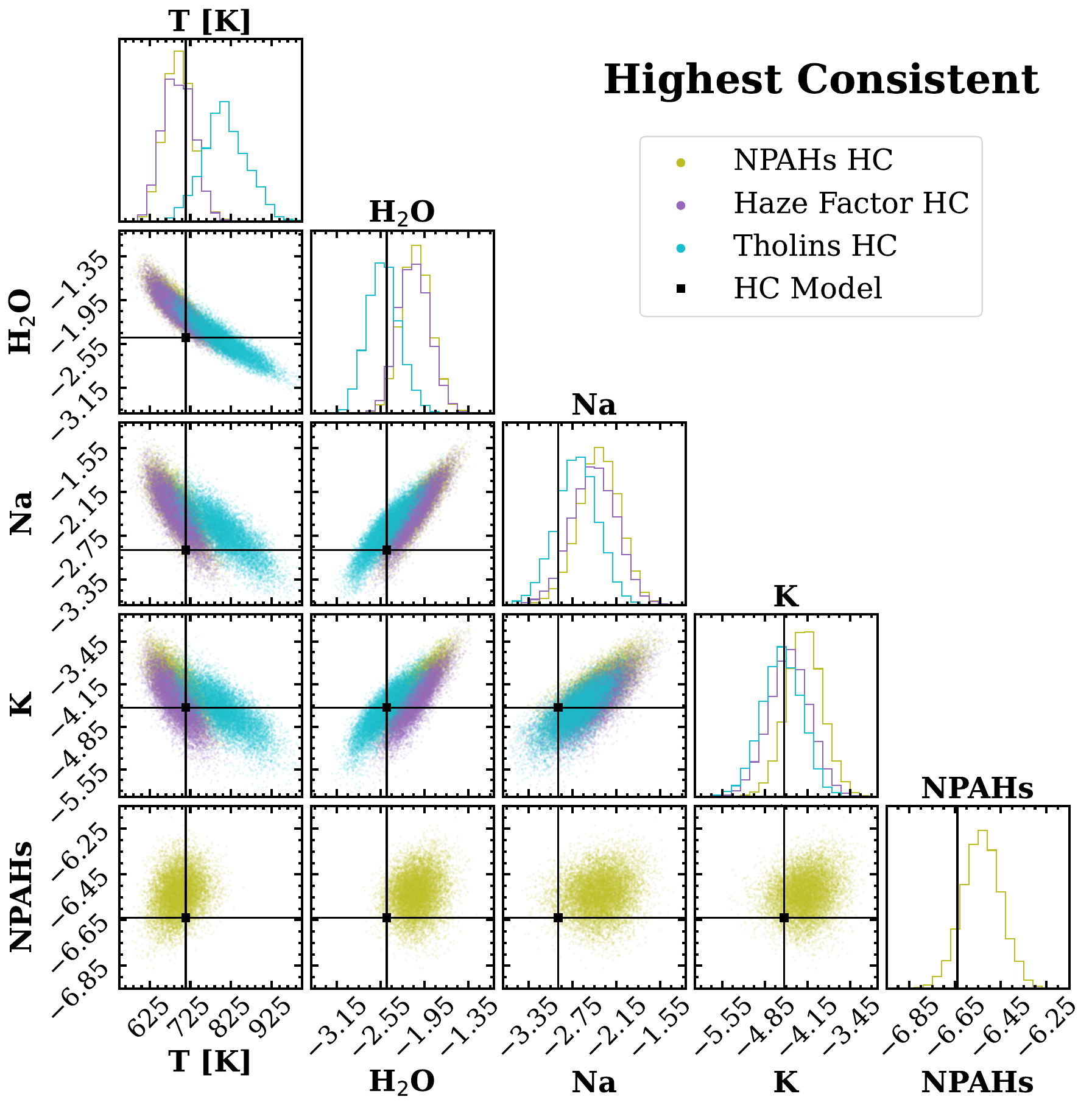}};
            \begin{scope}[x={(image.south east)},y={(image.north west)}]
            \node [anchor=north west] at (0,1) {(d)};
            \end{scope}
        \end{tikzpicture}
        \phantomcaption
        \label{fig:HighCont}
    \end{subfigure}  
    \caption{Results for the HL (a,c) and HC (b,d) retrieval runs. Subfigures (a) and (b): Top panels: Best-fitting spectra for the NPAH, Haze Factor and Tholin models (C20 plus PRISM data in black). Bottom panels: Residuals are plotted for each model in the respective colors showing the deviation of the data points from the spectra in units of the corresponding transit depth error ($\sigma$). The shaded bars indicate the 1, 3 and 5 $\sigma$ regions. Finally, the subpanel presents the model quality, showcasing the evidences as absolute deviations ($\Delta$ln$Z$) from the lower model (for details refer to Table~\ref{appendix:tab1}) and the reduced $\chi^2$.
    Subfigures (c) and (d): Corner plots including the temperature, H$_2$O, Na, K and PAH abundances. The 1D posteriors are presented as histograms on the diagonal and the 2D distributions are displayed by sample points. The input values for both the HL and HC models are given in black.}
    \label{CombRes}
\end{figure*}
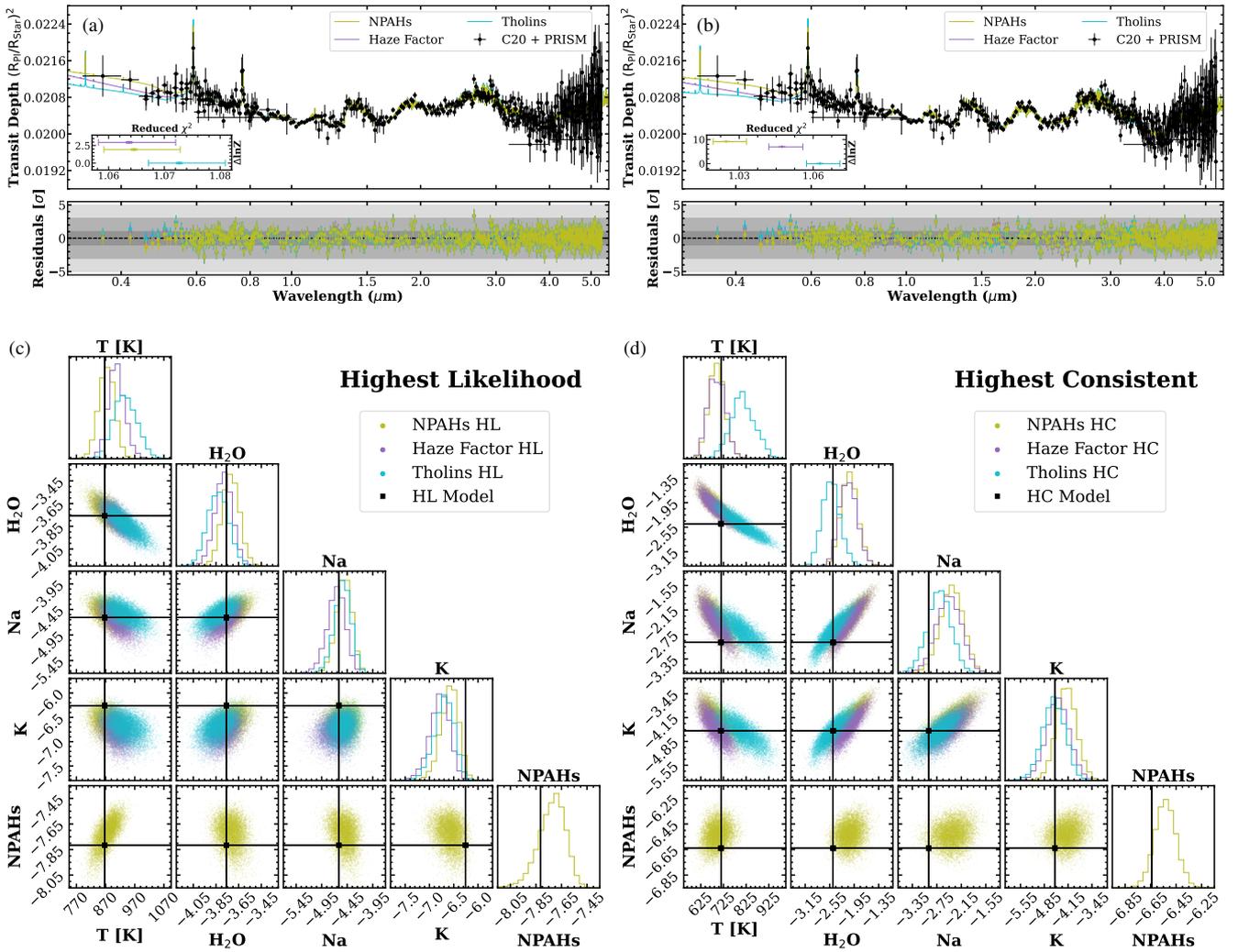
Following the results from section~\ref{sec: Carter}, we compute synthetic datasets for NIRSpec PRISM where we use the PAH model as an input for the official ETC PandExo 3.0 \citep{2017PASP..129f4501B}. As outlined below, we consider two cases with different PAH abundances to assess the detectability. The corresponding data are then added to the available set and further retrievals are conducted on the updated transmission spectra. Here, we compare the PAH model with the Haze Factor since, according to our previous findings, this synthetic slope source provides the largest improvements to the fit by adding the least amount of free parameters. Furthermore, we assess the performance of the second 'real' cloud/haze model, namely Tholins. Figure~\ref{CombRes} shows the best-fitting spectra and corner plots for both approaches. The full posteriors are again given in the appendix (ref. to Figures \ref{Full_PC_Post1} to \ref{Full_PC_Cloud_Post}).
\subsubsection{Highest likelihood abundance}
\label{Highest_Likelihood}
In the first case, we considered the highest likelihood solution from the original PAH retrieval in section~\ref{sec: Carter} since this model fits the spectrum best. According to Figure~\ref{PAH_MFs} and Table~\ref{appendix:tab2}, the corresponding PAH abundance measures $10^{-7.816}$. The resulting combined transmission spectrum, including both simulated and real data, is shown in Figure~\ref{fig:CP}. As mentioned, we conduct three retrievals on this dataset considering the PAH, Haze Factor and Tholin models with unchanged priors the results of which are again given in the Figure.

Accordingly, the best-fitting spectra show that by eye  each model can replicate the data quite well. This is at first surprising since in Figures~\ref{Carter_BF1} and~\ref{Carter_BF2} the overall shapes vary widely at wavelengths beyond 1.5 µm. The main differences between the models now appear in the optical regime with no significant offset around the 3.3 $\micron$ peak. This already indicates that the feature might be too weak to introduce a measurable effect. The $\chi^2$-evidence plot included in the sub-panel confirms this. Yet again, the reduced $\chi^2$ shows a substantial overlap between the PAH and Haze Factor models. Additionally, based on ln~$Z$ the Haze Factor is still slightly preferred despite the induced presence of PAHs in the atmosphere. Tholins, on the other hand, seem to be more disfavoured. Nevertheless, considering both metrics all models still remain feasible.

Therefore, the impact of the 3.3 $\micron$ feature seems negligible and thus even though the wavelength range now covers a significant part of the cross-sections the existence of PAHs cannot be verified. As the retrieval results in Table~\ref{appendix:tab2} show, a PAH abundance can in fact be retrieved. However, the Haze Factor still performs equally well. 
The corner plot in Figure~\ref{fig:HighLike} provides more insights. For the highest likelihood runs (HL), the temperatures of the Haze Factor and Tholin models and the PAH mass fractions are overestimated while the potassium mass fractions are slightly underestimated. Thus, there still seems to be a remaining degeneracy between the different free parameters. 

The retrieved PAH value of about $10^{-7.7}$ is also larger than the input which suggests that at these abundances PAHs could still not be detectable. Since the posterior drops off at around $10^{-7.5}$ this number could approach the true detection limit. Nevertheless, considering all results we can conclude, that the presence of PAHs on WASP-6~b at abundances of around $10^{-8}$ could most likely not be confirmed with NIRSpec PRISM. This situation could potentially change when observing multiple transits. Given the challenges in securing sufficient observation time for multiple transits with \textit{JWST}, we instead focus on a single transit.
\subsubsection{Highest consistent abundance}
\label{Highest_Consistent}
As shown above, the highest likelihood abundance would be in principal retrievable yet other models remain superior or equally likely. However, when reverting to the PAH mass fraction posterior in Figure~\ref{PAH_MFs} one can observe that the corresponding abundance is located near the lower boundary of the 1$\sigma$ intervals. As illustrated in the Figure, significantly higher mass fractions would however still be consistent with the statistics and might provide a stronger, detectable signal in the spectrum. Thus, in the following, we construct a synthetic dataset including a PAH abundance at $10^{-6.640}$ which is equal to the upper mass fraction boundary of the 1$\sigma$ region. For this, we fix PAHs to said value and run an additional retrieval on the C20 data to receive an optimized best-fitting spectrum. The details for this setup run are given in Tables~\ref{appendix:tab1} and~\ref{appendix:tab2} in the appendix. The resulting best-fitting spectrum is then used as an input for the ETC. The combined dataset is shown in Figure~\ref{fig:CPN}. 

As before, we conduct two retrievals considering the PAH, Haze Factor and Tholin models. 
The model quality results are shown in the subpanel of Figure~\ref{fig:CPN}. Accordingly, at this point, significant overlap is neither seen in the $\chi^2$ nor the evidences and the PAH model is clearly favoured. Thus, the Haze Factor and Tholin models are not able to replicate the data at a similar level anymore. Furthermore, we observe a slight positive offset of several Haze Factor and Tholin residuals around the 3.3 $\micron$ feature. This indicates that the influence of the peak on the transmission spectrum reached a measurable level. Nevertheless, the detection in this case is likely caused by the overall shape of the spectrum since this is not the only difference in the residual panel. An observation of PAHs based solely on the 3.3 $\micron$ peak would require higher abundances and/or more precise data. Additionally, the corner plot in Figure~\ref{fig:HighCont} shows that the accuracy does not increase with the higher PAH abundance. For example, the retrieved mass fractions for H$_2$O and Na are slightly overestimated. We suspect that the reason why this degeneracy remains lies in the random scatter of the simulated data. 

In conclusion, the results suggest that at a mass abundance of around $10^{-6.5}$ PAHs emerge as the favoured model and therefore could in theory be detected on WASP-6~b. 
\subsection{Discussion}

In the following section, we present a brief review of the implications of the results and limitations of our approach.
\subsubsection{Data Compatibility}
\label{sec:Offset}
\begin{figure}
    \includegraphics[width=\columnwidth]{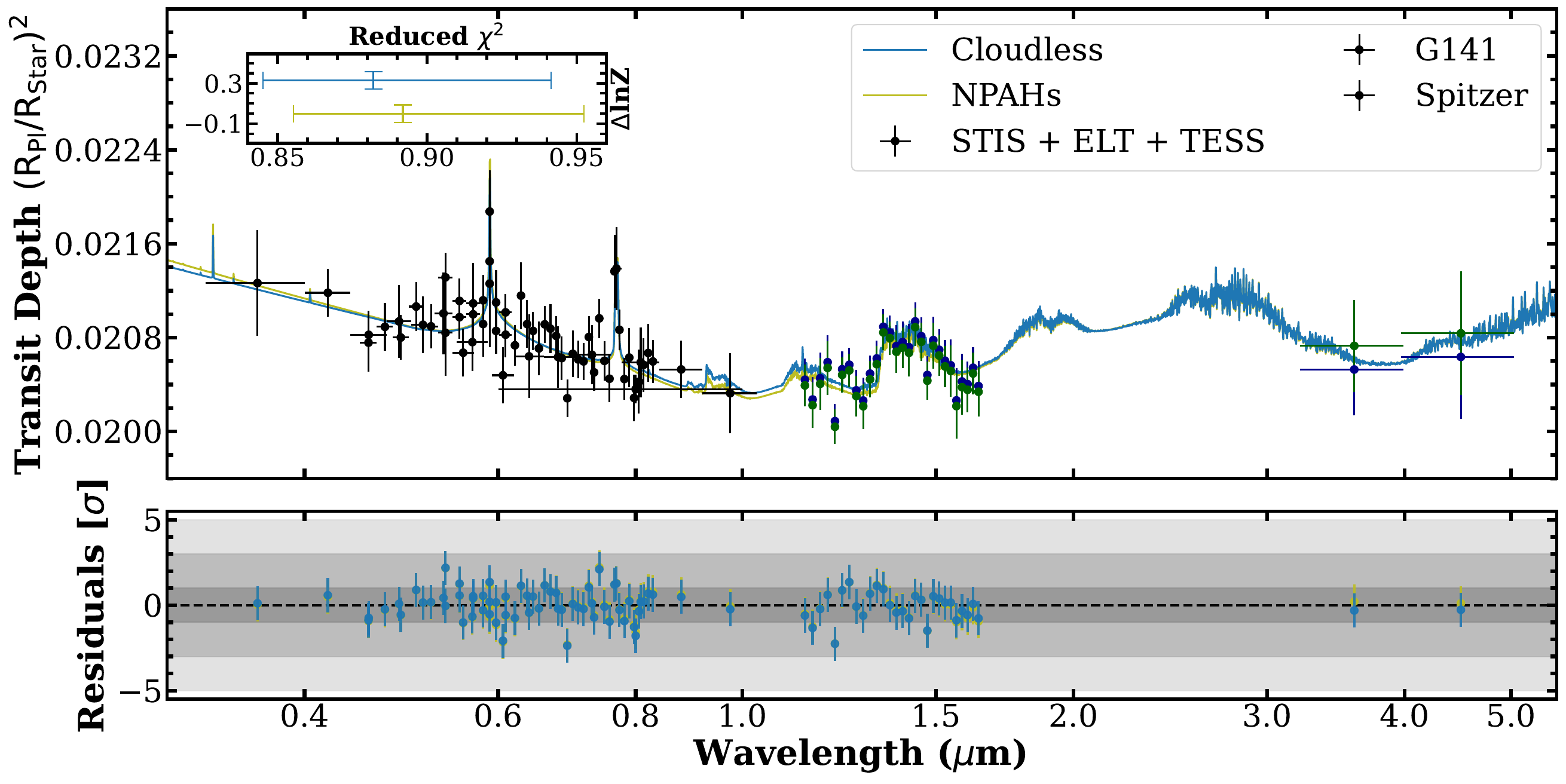}
    \caption{Results for the offset retrieval runs. Top panel: Best-fitting spectra for the Cloudless and PAH models (\textit{HST} STIS + \textit{ELT} + \textit{TESS} data in black, G141 and \textit{Spitzer} data including the best-fitting offsets for both retrievals in blue and green, respectively). Bottom panel: Residuals are plotted for each model in the respective colors showing the deviation of the data points from the spectra in units of the corresponding transit depth error ($\sigma$). The shaded bars indicate the 1, 3 and 5 $\sigma$ regions. Finally, the subpanel presents the model quality, showcasing the evidences as absolute deviations ($\Delta$ln$Z$) from the lower model (for details refer to Table~\ref{appendix:tab1}) and the reduced $\chi^2$.}
    \label{Offset}
\end{figure}

Throughout this work, we used a collection of data from \citet{2020MNRAS.494.5449C} as the basis of our analysis. Even though, in the mentioned study, the authors reanalysed the data from different sources using the same reduction techniques and corrected them for stellar variability, temporal variations or instrument limitations could render the individual sets incompatible. To test the impact on the results of such variations, we ran two additional retrievals including free data offsets. Here, we split the data into three individual sets based on the absence of an overlap. Thus, on one side the data from \textit{HST} STIS, \textit{ELT} and \textit{TESS} are combined into a single set, while on the other side, we construct uniform priors for the \textit{HST} G141 and \textit{Spitzer} data offsets with limits of 1000 and -2000 ppm (negative offsets shift the data upwards). We retrieved both the Cloudless and PAH model to assess the effect of the additionally introduced parameters. The results are summarized in Figure \ref{Offset}.

We find strong negative offsets for both models. Additionally, the model quality of both approaches is almost identical with the Cloudless scenario being slightly preferred. Therefore, as further displayed in Tables \ref{appendix:tab1} and \ref{appendix:tab2}, the inclusion of data offsets completely removes the PAH signal from the data. Nevertheless, these results can be explained under consideration of our previous findings. In section \ref{sec: Carter}, we observed strong CIA in the cases of the slope-less models which was caused by the need for significant Rayleigh scattering. Because of that, the best-fitting spectra deviate notably from the final points of the \textit{Hubble} G141 data as well as the \textit{Spitzer} observations which are better fit by the cloud/haze approaches (see Figures \ref{Carter_BF1} and \ref{Carter_BF2}). Therefore, by introducing these offsets, the need for slope sources is removed and the presence of the prominent CIA feature encouraged. Since, as mentioned, the latter has never been previously observed, we excluded the possibilities of offsets completely from our previous work and instead analyzed the data in its current state without bringing in any additional biases. Nevertheless, this increases the value of NIRSpec PRISM as an instrument mode since its wavelength coverage overlaps with each of the three split datasets. Therefore, future observations would reveal any inconsistencies within the C20 data.

We further acknowledge that, in real observations, potential inconsistencies may arise between the C20 and PRISM datasets. While detecting PAHs using only NIRSpec PRISM would require higher abundances in the 2$\sigma$ range we believe a combined analysis remains feasible, with a detection limit close to the value indicated in this work.
\subsubsection{Reference ISM values}
In section~\ref{sec:PRISM}, we found that a relative PAH mass abundance of around $10^{-6.5}$ could be detectable with NIRSpec PRISM in contrast to cases with a fraction of $10^{-8}$ for which a detection would be challenging. To put this into context, we can compare these values with the ISM abundance mentioned before. According to \citet{2022MNRAS.512..430E}, a mass fraction of $10^{-7}$ equals approximately 1/1000 of the ISM value. Since both studies describe hydrogen and helium-dominated atmospheres, we can use that number to get an order-of-magnitude estimate. Therefore, while 0.01 per cent of the ISM value appears to be not detectable, a value of around 0.3 per cent could be. We additionally observed in section~\ref{Highest_Likelihood} that an abundance of $10^{-7.5}$ or about 0.03 per cent possibly won't introduce a difference large enough to be detectable either. Therefore, we are likely left with the detection limit located in between at approximately 0.1 per cent. Thus, this work infact confirms the claims from \citet{2022MNRAS.512..430E} that this mass fraction could be detectable although, under the consideration of several contributing molecules, the 3.3 $\micron$ does not remain as pronounced as forecasted. In conclusion, WASP-6~b presents a promising target for a future PAH detection, assuming that at least 0.1 per cent of the ISM abundance survives the planet formation process or is replenished a posteriori. 

We note that this detection limit holds exclusively for the case of WASP-6 b. However, a significant portion of the giant exoplanets known today are located closer to Earth. Thus, considering the increased precision one is able to achieve on brighter targets, the cutoff might decrease slightly. Nevertheless, in this work, we observed that the PAH signature is significantly masked at lower abundances and thus, we do not expect the limit to change drastically.

\subsubsection{Dectectability dependence on other molecules}
\label{Mol}
The currently available data shows features of only three line species namely sodium, potassium and water which were already confirmed in C20. Nevertheless, other exoplanet observations for example of WASP-39 b \citep{2023Natur.614..659R} show large features from other molecules like CO$_2$, SO$_2$, CO, CH$_4$, etc. some of which were included in our simulations yet no signal could be retrieved. Thus, an observation with NIRSpec PRISM would likely also reveal additional molecules in the wavelength regime covered in our retrievals. Additional molecules could introduce new features at longer wavelengths and potentially overshadow the 3.3 $\micron$ PAH peak. As an example other carbon-bearing species like methane also produce peaks around these wavelengths \citep[see e.g. ExoMol][]{2017A&A...605A..95Y}. Nevertheless, these features tend to be narrower and thus potentially distinct from the broader 3.3 $\micron$ feature produced by PAHs. In fact, our retrievals on the combined datasets already indicate that despite the presence of the 3.3 $\micron$ peak, we do not detect any robust CH$_4$ signal. Moreover, since the PAH cross-sections contain more features in the mid-infrared regime a potential degeneracy could be lifted with other instruments like for example \textit{JWST} MIRI. 

In conclusion, we do not expect the detection limit for PAH abundances to change significantly with more information on additional molecules because of the global influence of the cross-sections on the transmission spectrum. To quantify this effect and further narrow down the detection limit a complete prior analysis would be necessary including a multitude of major line species which is outside of the scope of this work.

\section{Summary}
\label{Conclusion}
In this study, we performed atmospheric retrievals on the currently available dataset for WASP-6~b \citep{2020MNRAS.494.5449C} in order to assess if these molecules can be detected with \textit{James Webb} in the future. To this aim, we compared different haze and cloud opacities and their impact on the transmission spectrum of this giant planet. 

We show that current observations are consistent with PAHs being present in the cloud structure of WASP-6~b. While we could not rule out alternative haze models with the existing dataset, we show that observations with NIRSpec PRISM that additionally cover the 3.3 $\micron$ feature could be able to detect PAHs at an abundance as low as 0.1 per cent of the ISM value (mass fraction of approximately~$10^{-7}$). 

Based on these results, our study strongly motivates observational campaigns with \textit{JWST} NIRSpec PRISM of the hot Saturn WASP-6~b in order to constrain the abundance of PAHs in its atmosphere and possibly detect these molecules for the first time on exoplanets. 

\section*{Acknowledgements}
We acknowledge support of the DFG (German Research Foundation), Research Unit “Transition discs” - 325594231 and of the Excellence Cluster ORIGINS - EXC-2094 - 390783311. The simulations have been carried out on the computing facilities of the Computational Center for Particle and Astrophysics (C2PAP).\\ \\
This research has made use of the NASA Exoplanet Archive, which is operated by the California Institute of Technology, under contract with the National Aeronautics and Space Administration under the Exoplanet Exploration Program.
\\ \\
This publication makes use of data products from the Two Micron All Sky Survey, which is a joint project of the University of Massachusetts and the Infrared Processing and Analysis Center/California Institute of Technology, funded by the National Aeronautics and Space Administration and the National Science Foundation.\\ \\
We thank the reviewer for their valuable feedback and constructive suggestions which have greatly improved the quality of our work.\\ \\
\textit{Softwares}: Our Python codes use the following libraries: \texttt{Numpy} \citep{2020Natur.585..357H}, \texttt{Matplotlib} \citep{2007CSE.....9...90H}, \texttt{Jupyter} \citep{2016ppap.book...87K}, \texttt{Dill} \citep{2012arXiv1202.1056M,pathos-framework}, and \texttt{Seaborn} \citep{2021JOSS....6.3021W}.

%%%%%%%%%%%%%%%%%%%%%%%%%%%%%%%%%%%%%%%%%%%%%%%%%%
\section*{Data Availability}

The data underlying this article are available from the authors upon request.



\bibliographystyle{mnras}
\bibliography{ref} % if your bibtex file is called example.bib

%\begin{thebibliography}{99}
%\end{thebibliography}

%%%%%%%%%%%%%%%%%%%%%%%%%%%%%%%%%%%%%%%%%%%%%%%%%%

%%%%%%%%%%%%%%%%% APPENDICES %%%%%%%%%%%%%%%%%%%%%
\newpage
\appendix

\section{Retrieval Results}
\label{appendix}
\begin{figure*}
    \centering
    \includegraphics[width=0.85\textwidth]{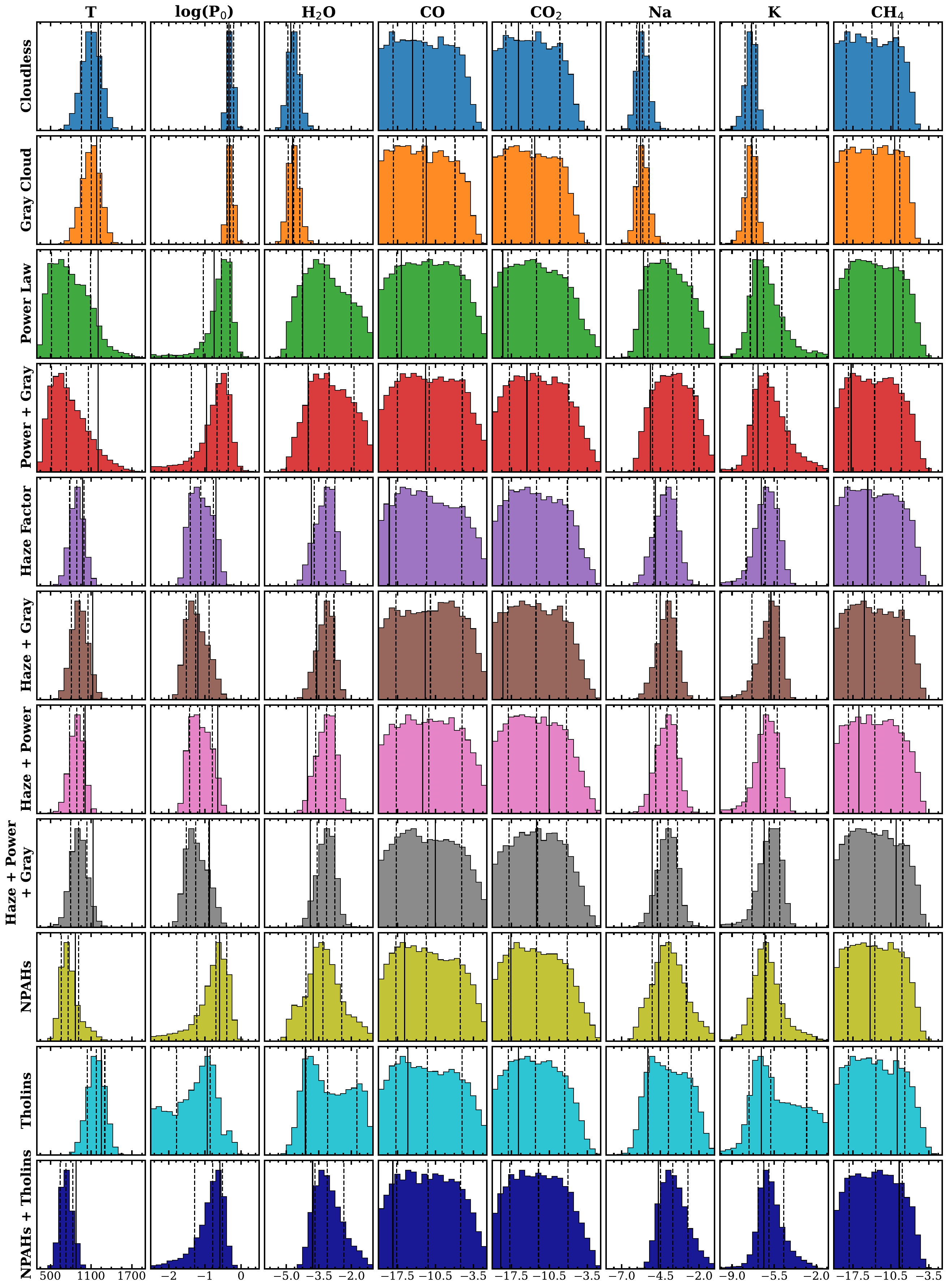}
    \caption{Results for the CD retrievals (refer to Table \ref{tab:Setups}). Displayed are the 1D posteriors of the base parameters shared by every model: $T$[K], log($P_\mathrm{0}$)~[log(bar)], H$_2$O, CO, CO$_2$, Na, K and CH$_4$. The medians as well as the 16 and 84 percentile limits are given by the dashed lines in black. Furthermore, black solid lines indicate the highest-likelihood values.}
    \label{Full_Carter_Post}
\end{figure*}

\begin{figure*}
    \centering
    \includegraphics[width=0.7\textwidth]{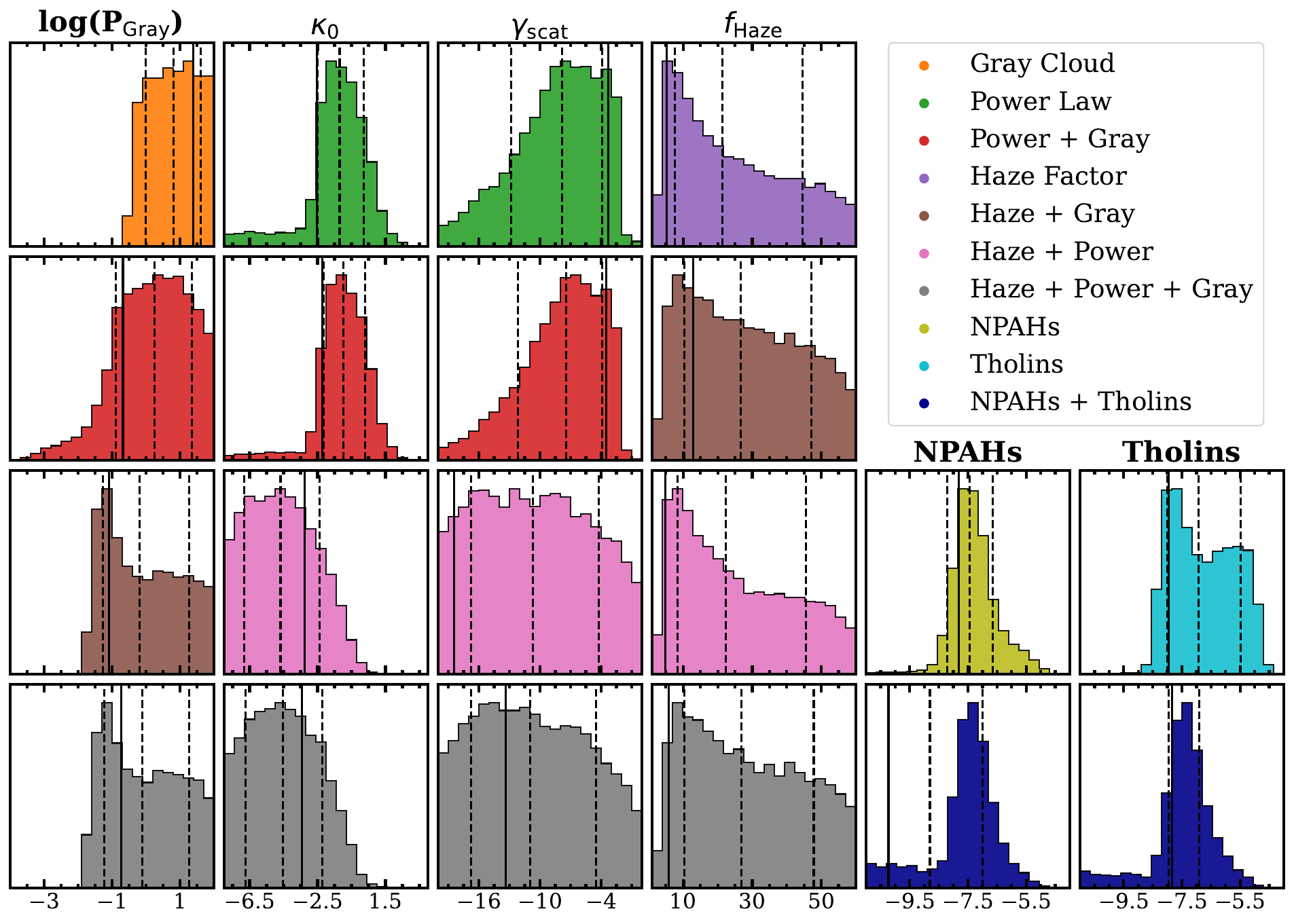}
    \caption{Results for the CD retrievals (refer to Table \ref{tab:Setups}). Displayed are the 1D posteriors of the cloud parameters that distinguish the models from each other: log($P_{\mathrm{Gray}}$)~[log(bar)], $\kappa_\mathrm{0}$~[log(cm$^2$/g)], $\gamma_{\mathrm{scat}}$, $f_{\mathrm{Haze}}$, PAHs, and Tholins. The medians as well as the 16 and 84 percentile limits are given by the dashed lines in black. Furthermore, black solid lines indicate the highest-likelihood values.}
    \label{Full_Carter_Cloud_Post}
\end{figure*}
\begin{figure*}
    \centering
    \includegraphics[width=0.85\textwidth]{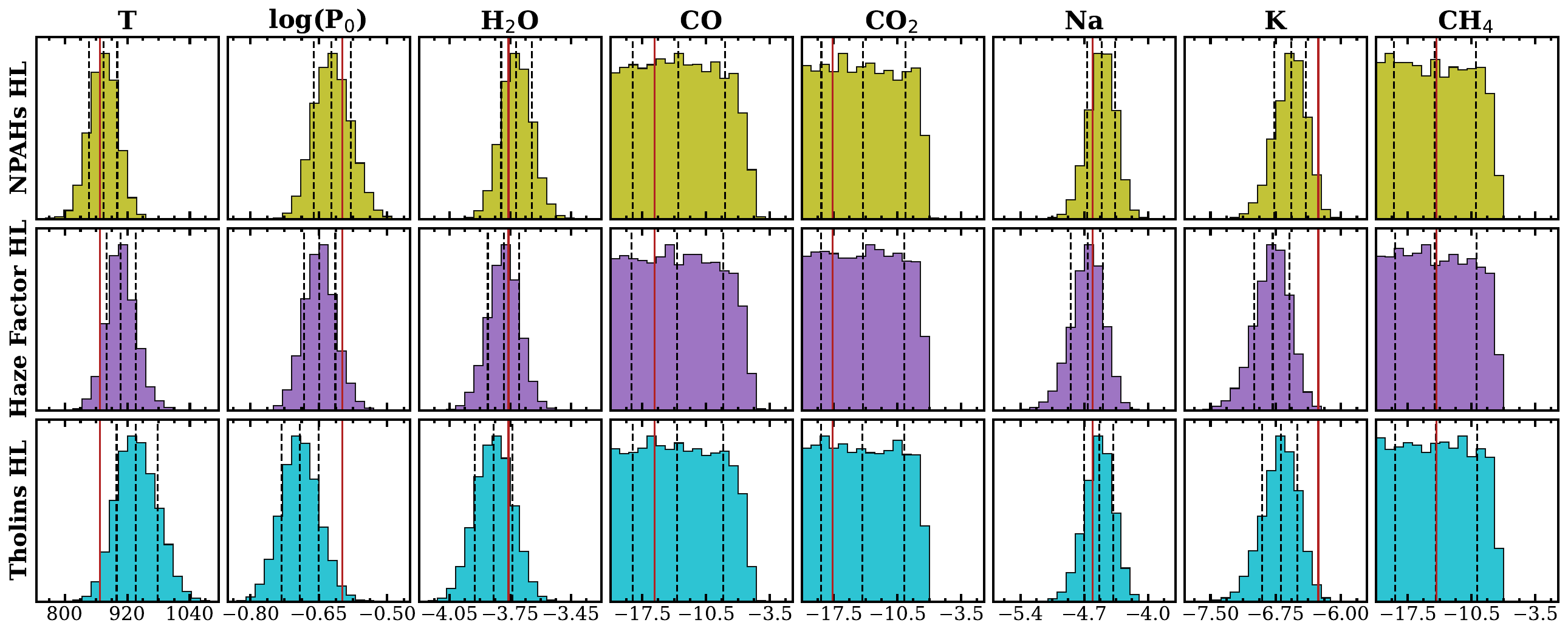}
    \caption{Results for the HL retrievals (refer to Table \ref{tab:Setups}). Displayed are the 1D posteriors of the base parameters shared by every model: $T$[K], log($P_\mathrm{0}$)~[log(bar)], H$_2$O, CO, CO$_2$, Na, K and CH$_4$. The medians as well as the 16 and 84 percentile limits are given by the dashed lines in black. Furthermore, red solid lines indicate the input values from the respective PAH retrieval on the C20 data (refer to section \ref{sec:PRISM} for more details).}
    \label{Full_PC_Post1}
\end{figure*}
\begin{figure*}
    \centering
    \includegraphics[width=0.85\textwidth]{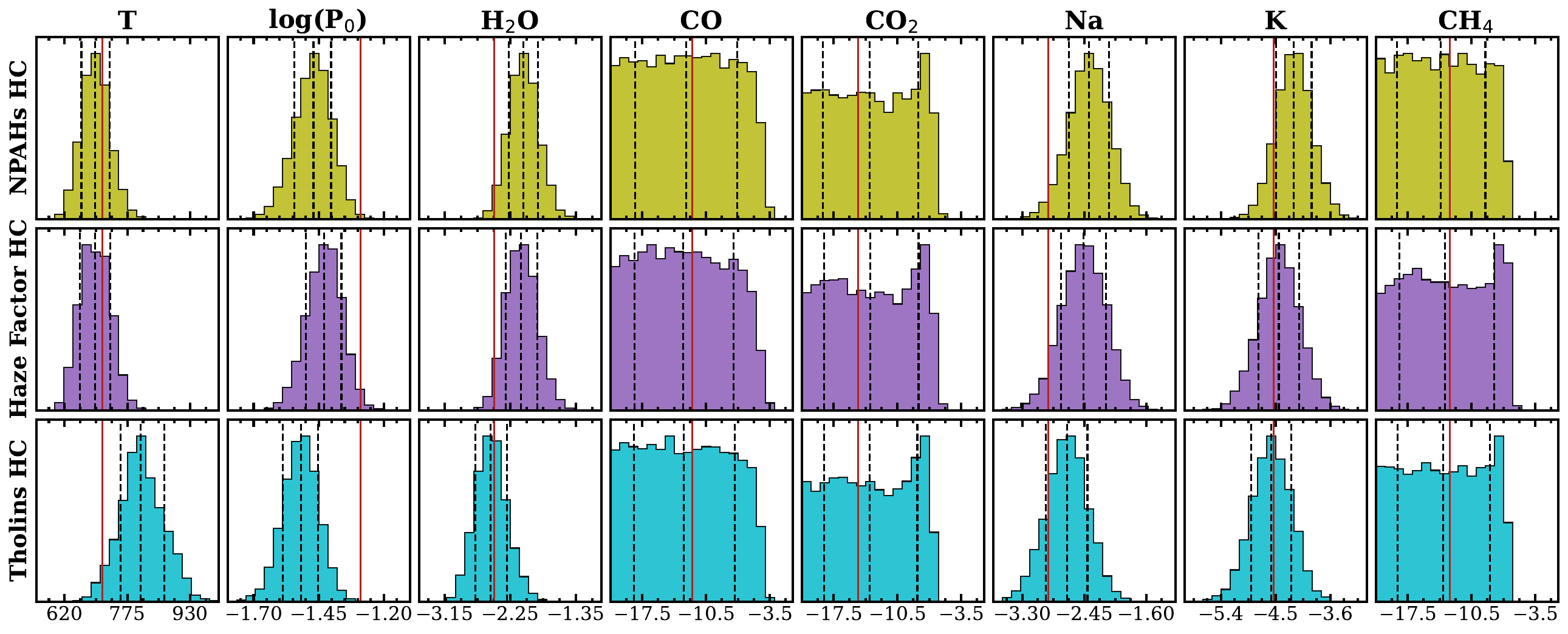}
    \caption{Results for the HC retrievals (refer to Table \ref{tab:Setups}). Displayed are the 1D posteriors of the base parameters shared by every model: $T$[K], log($P_\mathrm{0}$)~[log(bar)], H$_2$O, CO, CO$_2$, Na, K and CH$_4$. The medians as well as the 16 and 84 percentile limits are given by the dashed lines in black. Furthermore, red solid lines indicate the input values from the respective PAH retrieval on the C20 data (refer to section \ref{sec:PRISM} for more details).}
    \label{Full_PC_Post2}
\end{figure*}
\begin{figure*}
    \centering
    \includegraphics[width=0.7\textwidth]{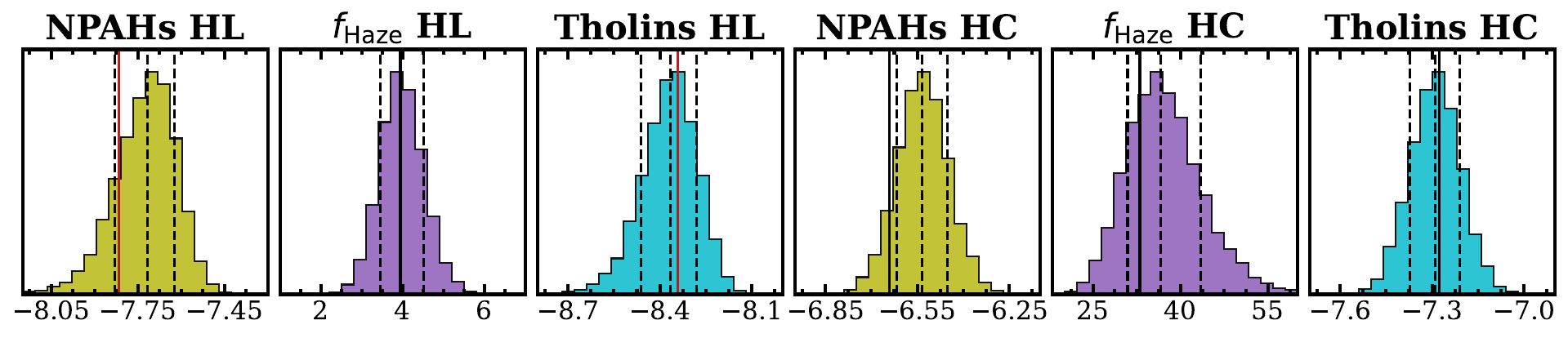}
    \caption{Results for the HL and HC retrievals (refer to Table \ref{tab:Setups}). Displayed are the 1D posteriors of the cloud parameters that distinguish the models from each other: PAHs, $f_{\mathrm{Haze}}$ and  Tholins. The medians as well as the 16 and 84 percentile limits are given by the dashed lines in black. Furthermore, red and black solid lines indicate the input PAH values from the respective retrieval on the C20 data (refer to section \ref{sec:PRISM} for more details) as well as the highest-likelihood Haze Factors and Tholin opacities.}
    \label{Full_PC_Cloud_Post}
\end{figure*}

\begin{table}
\centering
\rotatebox{90}{%
\begin{minipage}{0.9\textheight}
\def\arraystretch{1}
\tiny\centering
\setlength{\tabcolsep}{1.4mm}
\scriptsize
\centering
{\fontsize{6.3}{8}\selectfont 
\begin{tabular}{l*{12}{c}}
Retrieval & ln~$Z$ & $\chi^2$ & $T$[K]&log($P_\mathrm{0}$) [log(bar)]&H$_2$O&CO&CO$_2$&Na&K&CH$_4$\\
\hline
\hline
\multicolumn{11}{c}{\centering \citet{2020MNRAS.494.5449C}} \\
\hline
Cloudless & 652.6 $\pm$ 0.1 & 0.91 & 1100 $\pm$ 142 (1206) & -0.299 $\pm$ 0.086 (-0.3353) & -4.67 $\pm$ 0.28 (-4.799) & -12.7 $\pm$ 5.7 (-14.7) & -13.6 $\pm$ 5.0 (-16.19) & -5.66 $\pm$ 0.40 (-5.855) & -7.42 $\pm$ 0.46 (-7.369) & -14.0 $\pm$ 4.8 (-10.13) \\
Gray Cloud & 651.5 $\pm$ 0.1 & 0.92 & 1101 $\pm$ 140 (1182) & -0.304 $\pm$ 0.085 (-0.3257) & -4.66 $\pm$ 0.27 (-4.716) & -12.8 $\pm$ 5.7 (-12.2) & -13.7 $\pm$ 5.1 (-13.18) & -5.64 $\pm$ 0.40 (-5.771) & -7.40 $\pm$ 0.46 (-7.358) & -13.7 $\pm$ 4.9 (-9.777) \\
Power Law Cloud & 654.4 $\pm$ 0.1 & 0.84 & 768 $\pm$ 288 (1205) & -0.54 $\pm$ 0.37 (-0.7445) & -3.2 $\pm$ 1.1 (-4.247) & -11.7 $\pm$ 6.0 (-16.76) & -12.8 $\pm$ 5.5 (-19.08) & -4.0 $\pm$ 1.4 (-5.591) & -6.4 $\pm$ 1.3 (-6.915) & -13.4 $\pm$ 5.0 (-10.06) \\
Power + Gray & 653.5 $\pm$ 0.1 & 0.84 & 734 $\pm$ 274 (1203) & -0.66 $\pm$ 0.52 (-0.9524) & -3.0 $\pm$ 1.1 (-3.99) & -11.6 $\pm$ 6.0 (-12.33) & -12.5 $\pm$ 5.5 (-14.61) & -3.7 $\pm$ 1.4 (-5.154) & -6.1 $\pm$ 1.4 (-6.839) & -13.5 $\pm$ 4.9 (-17.81) \\
Haze Factor & 656.1 $\pm$ 0.1 & 0.83 & 891 $\pm$ 107 (972) & -1.12 $\pm$ 0.32 (-0.7006) & -3.17 $\pm$ 0.48 (-3.85) & -12.1 $\pm$ 6.1 (-19.01) & -12.9 $\pm$ 5.4 (-19.04) & -4.12 $\pm$ 0.74 (-4.817) & -6.2 $\pm$ 1.3 (-6.56) & -13.6 $\pm$ 5.1 (-14.74) \\
Haze + Gray & 655.8 $\pm$ 0.1 & 0.83 & 927 $\pm$ 122 (1123) & -1.25 $\pm$ 0.32 (-1.194) & -3.16 $\pm$ 0.41 (-3.588) & -11.4 $\pm$ 6.2 (-12.37) & -12.9 $\pm$ 5.4 (-19.03) & -4.04 $\pm$ 0.65 (-4.511) & -5.9 $\pm$ 1.2 (-5.773) & -13.5 $\pm$ 5.0 (-15.4) \\
Haze + Power & 656 $\pm$ 0.1 & 0.85 & 890 $\pm$ 106 (1011) & -1.14 $\pm$ 0.31 (-0.6469) & -3.14 $\pm$ 0.45 (-4.025) & -11.8 $\pm$ 6.1 (-12.86) & -12.8 $\pm$ 5.4 (-10.52) & -4.07 $\pm$ 0.70 (-5.207) & -6.2 $\pm$ 1.3 (-6.669) & -13.5 $\pm$ 5.0 (-16.37) \\
Haze + Power + Gray & 655.7 $\pm$ 0.1 & 0.85 & 916 $\pm$ 122 (1129) & -1.26 $\pm$ 0.31 (-0.8809) & -3.14 $\pm$ 0.41 (-3.899) & -11.9 $\pm$ 6.1 (-10.51) & -12.7 $\pm$ 5.4 (-12.86) & -4.00 $\pm$ 0.65 (-5.034) & -5.9 $\pm$ 1.2 (-6.319) & -13.4 $\pm$ 5.0 (-9.501) \\
NPAHs & 653.7 $\pm$ 0.1 & 0.85 & 761 $\pm$ 130 (867.3) & -0.70 $\pm$ 0.42 (-0.5984) & -3.31 $\pm$ 0.83 (-3.758) & -12.2 $\pm$ 6.0 (-16.15) & -12.8 $\pm$ 5.5 (-17.6) & -4.0 $\pm$ 1.1 (-4.61) & -6.2 $\pm$ 1.2 (-6.257) & -13.4 $\pm$ 5.0 (-14.33) \\
Tholins & 655.1 $\pm$ 0.1 & 0.82 & 1174 $\pm$ 130 (1251) & -1.8 $\pm$ 1.2 (-0.9372) & -3.1 $\pm$ 1.2 (-4.118) & -11.9 $\pm$ 6.1 (-15.56) & -13.1 $\pm$ 5.2 (-16.17) & -4.0 $\pm$ 1.4 (-5.305) & -5.8 $\pm$ 2.4 (-6.577) & -13.3 $\pm$ 5.3 (-9.287) \\
NPAHs + Tholins & 653.7 $\pm$ 0.1 & 0.85 & 728 $\pm$ 93 (876.1) & -0.79 $\pm$ 0.39 (-0.5847) & -3.11 $\pm$ 0.68 (-3.786) & -11.9 $\pm$ 5.9 (-18.34) & -12.5 $\pm$ 5.4 (-19.43) & -3.70 $\pm$ 0.91 (-4.64) & -5.9 $\pm$ 1.1 (-6.36) & -13.3 $\pm$ 4.9 (-8.946) \\
\hline
\multicolumn{11}{c}{\citet{2020MNRAS.494.5449C} + NIRSpec PRISM Highest Likelihood} \\
\hline
NPAHs & 3558.6 $\pm$ 0.1 & 1.05 & 874 $\pm$ 27 (876) & -0.622 $\pm$ 0.040 (-0.6258) & -3.719 $\pm$ 0.076 (-3.731) & -13.6 $\pm$ 5.1 (-13.06) & -14.3 $\pm$ 4.6 (-17.48) & -4.51 $\pm$ 0.15 (-4.54) & -6.57 $\pm$ 0.18 (-6.58) & -14.6 $\pm$ 4.5 (-15.93) \\
Haze Factor & 3559.6 $\pm$ 0.1 & 1.05 & 907 $\pm$ 28 (909.9) & -0.648 $\pm$ 0.035 (-0.6564) & -3.782 $\pm$ 0.078 (-3.769) & -13.7 $\pm$ 5.0 (-12.77) & -14.3 $\pm$ 4.6 (-12.48) & -4.66 $\pm$ 0.18 (-4.637) & -6.78 $\pm$ 0.20 (-6.76) & -14.5 $\pm$ 4.5 (-14.24) \\
Tholins & 3556.7 $\pm$ 0.1 & 1.06 & 937 $\pm$ 40 (946.7) & -0.692 $\pm$ 0.041 (-0.7037) & -3.830 $\pm$ 0.093 (-3.831) & -13.7 $\pm$ 5.0 (-14.45) & -14.4 $\pm$ 4.6 (-9.763) & -4.54 $\pm$ 0.16 (-4.528) & -6.69 $\pm$ 0.20 (-6.69) & -14.4 $\pm$ 4.5 (-19.7) \\ 
\hline
\multicolumn{11}{c}{Highest Consistent (Setup Run)} \\ 
\hline
NPAHs & 655.4 $\pm$ 0.1 & 0.87 & 683 $\pm$ 74 (713.9) & -1.21 $\pm$ 0.15 (-1.289) & -2.52 $\pm$ 0.30 (-2.465) & -11.4 $\pm$ 6.3 (-12.01) & -11.7 $\pm$ 6.0 (-14.78) & -2.94 $\pm$ 0.45 (-2.946) & -4.96 $\pm$ 0.81 (-4.533) & -13.2 $\pm$ 5.2 (-12.85) \\
\hline
\multicolumn{11}{c}{\citet{2020MNRAS.494.5449C} + NIRSpec PRISM Highest Consistent} \\
\hline
NPAHs & 3569.4 $\pm$ 0.1 & 1.01 & 696 $\pm$ 35 (708.3) & -1.470 $\pm$ 0.072 (-1.464) & -2.07 $\pm$ 0.20 (-2.129) & -12.7 $\pm$ 5.6 (-16.19) & -13.5 $\pm$ 5.3 (-7.09) & -2.38 $\pm$ 0.28 (-2.468) & -4.20 $\pm$ 0.30 (-4.275) & -13.9 $\pm$ 4.9 (-12.39) \\
Haze Factor & 3567.2 $\pm$ 0.1 & 1.03 & 695 $\pm$ 38 (711.4) & -1.429 $\pm$ 0.069 (-1.372) & -2.10 $\pm$ 0.22 (-2.259) & -13.0 $\pm$ 5.4 (-17.02) & -13.4 $\pm$ 5.2 (-7.022) & -2.46 $\pm$ 0.31 (-2.751) & -4.44 $\pm$ 0.34 (-4.602) & -13.4 $\pm$ 5.3 (-6.932) \\
Tholins & 3560.3 $\pm$ 0.1 & 1.05 & 808 $\pm$ 55 (802.7) & -1.519 $\pm$ 0.068 (-1.518) & -2.52 $\pm$ 0.22 (-2.493) & -12.9 $\pm$ 5.6 (-14.08) & -13.5 $\pm$ 5.1 (-7.431) & -2.69 $\pm$ 0.29 (-2.711) & -4.57 $\pm$ 0.33 (-4.496) & -13.6 $\pm$ 5.1 (-14.84) \\
\hline
\multicolumn{11}{c}{\citet{2020MNRAS.494.5449C} Offset Analysis} \\
\hline
Cloudless & 655.6 $\pm$ 0.1 & 0.81 & 1025 $\pm$ 140 (1022) & -0.03 $\pm$ 0.15 (-0.06294) & -4.30 $\pm$ 0.35 (-4.259) & -11.9 $\pm$ 6.2 (-18.4) & -12.8 $\pm$ 5.7 (-17.8) & -6.33 $\pm$ 0.50 (-6.21) & -8.18 $\pm$ 0.69 (-7.908) & -13.4 $\pm$ 5.3 (-9.968) \\
NPAHs & 655.3 $\pm$ 0.1 & 0.82 & 1008 $\pm$ 145 (1118) & -0.04 $\pm$ 0.15 (-0.1535) & -4.26 $\pm$ 0.38 (-4.486) & -11.8 $\pm$ 6.1 (-15.6) & -12.5 $\pm$ 5.7 (-16.13) & -6.28 $\pm$ 0.55 (-6.162) & -8.13 $\pm$ 0.72 (-7.957) & -13.3 $\pm$ 5.3 (-20.42) \\

\hline
\hline
\end{tabular}%
}
\end{minipage}}
\caption{Retrievel results for all models divided into the different datasets and approaches. Featured Parameters: ln~$Z$, best-fitting reduced $\chi^2$, $T$[K], log($P_\mathrm{0}$)~[log(bar)], H$_2$O, CO, CO$_2$, Na, K, and CH$_4$.}
\label{appendix:tab1}  

\end{table}

\begin{table}

\centering

\rotatebox{90}{%
\begin{minipage}{0.9\textheight}
\def\arraystretch{1}
\tiny\centering
\setlength{\tabcolsep}{1.4mm}
\scriptsize
\centering
{\fontsize{6.3}{8}\selectfont 
\begin{tabular}{l*{9}{c}}
Retrieval&log($P_{\mathrm{Gray}}$) [log(bar)]&  $\kappa_\mathrm{0}$ [log(cm$^2$/g)]&$\gamma_{\mathrm{scat}}$&$f_{\mathrm{Haze}}$ &NPAHs&Tholins&$\Delta_{\mathrm{G141}}$&$\Delta_{\mathrm{\textit{Spitzer}}}$\\
\hline
\hline
\multicolumn{8}{c}{\centering \citet{2020MNRAS.494.5449C}} \\
\hline
Cloudless & - & - & - & - & - & - & - & - \\
Gray Cloud & 0.81 $\pm$ 0.81 (1.395) & - & - & - & - & - & - & - \\
Power Law Cloud & - & -1.2 $\pm$ 1.4 (-2.53) & -7.8 $\pm$ 4.5 (-3.285) & - & - & - & - & - \\
Power + Gray & 0.3 $\pm$ 1.1 (-0.6794) & -1.0 $\pm$ 1.2 (-2.216) & -7.4 $\pm$ 4.2 (-3.484) & - & - & - & - & - \\
Haze Factor & - & - & - & 21 $\pm$ 19 (5.288) & - & - & - & - \\
Haze + Gray & -0.2 $\pm$ 1.3 (-1.096) & - & - & 27 $\pm$ 19 (12.98) & - & - & - & - \\
Haze + Power & - & -4.7 $\pm$ 2.2 (-3.262) & -10.7 $\pm$ 6.3 (-18.44) & 22 $\pm$ 19 (4.83) & - & - & - & - \\
Haze + Power + Gray & -0.1 $\pm$ 1.3 (-0.7305) & -4.5 $\pm$ 2.3 (-3.419) & -11.0 $\pm$ 6.2 (-13.38) & 27 $\pm$ 19 (5.821) & - & - & - & - \\
NPAHs & - & - & - & - & -7.44 $\pm$ 0.79 (-7.816) & - & - & - \\
Tholins & - & - & - & - & - & -6.9 $\pm$ 1.3 (-7.958) & - & - \\
NPAHs + Tholins & - & - & - & - & -8.7 $\pm$ 5.0 (-10.23) & -8.0 $\pm$ 4.9 (-7.84) & - & - \\
\hline
\multicolumn{8}{c}{\citet{2020MNRAS.494.5449C} + NIRSpec PRISM Highest Likelihood} \\
\hline
NPAHs & - & - & - & - & -7.72 $\pm$ 0.10 (-7.685) & - & - & - \\
Haze Factor & - & - & - & 3.97 $\pm$ 0.54 (3.949) & - & - & - & - \\
Tholins & - & - & - & - & - & -8.366 $\pm$ 0.091 (-8.342) & - & - \\
\hline
\multicolumn{8}{c}{Highest Consistent (Setup Run)} \\
\hline
NPAHs & - & - & - & - & -6.640 (fixed) & - & - & - \\
\hline
\multicolumn{8}{c}{\citet{2020MNRAS.494.5449C} + NIRSpec PRISM Highest Consistent} \\
\hline
NPAHs & - & - & - & - & -6.534 $\pm$ 0.083 (-6.536) & - & - & - \\
Haze Factor & - & - & - & 36.7 $\pm$ 6.3 (33.04) & - & - & - & - \\ 
Tholins & - & - & - & - & - & -7.290 $\pm$ 0.082 (-7.276) & - & - \\
\hline
\multicolumn{8}{c}{\citet{2020MNRAS.494.5449C} Offset Analysis} \\
\hline
Cloudless & - & - & - & - & - & - & -0.000208 $\pm$ 0.000080 (-0.0001844) & -0.00094 $\pm$ 0.00032 (-0.000759) \\
NPAHs & - & - & - & - & -14.6 $\pm$ 4.3 (-20.82) & - & -0.000201 $\pm$ 0.000084 (-0.0001358) & -0.00093 $\pm$ 0.00033 (-0.0009603) \\
\hline
\hline
\end{tabular}%
}
\end{minipage}}
\caption{Retrievel results for all models divided into the different datasets and approaches. Featured Parameters: log($P_{\mathrm{Gray}}$)~[log(bar)], $f_{\mathrm{Haze}}$, $\kappa_\mathrm{0}$~[log(cm$^2$/g)], $\gamma_{\mathrm{scat}}$, PAHs, Tholins, $\Delta_{\mathrm{G141}}$, and $\Delta_{\mathrm{\textit{Spitzer}}}$.}
\label{appendix:tab2}  

\end{table}

%%%%%%%%%%%%%%%%%%%%%%%%%%%%%%%%%%%%%%%%%%%%%%%%%%

% Don't change these lines
\bsp	% typesetting comment
\label{lastpage}
\end{document}